\documentclass[ASNA,twocolumn]{USG}

\usepackage{anyfontsize}
\usepackage{chemformula}

\graphicspath{{./Figure/}}

\articletype{RESEARCH ARTICLE}
\startpage{1}

\makeatletter
\historydates{}
\patchcmd{\@maketitle}{\includegraphics[width=26.5mm]{Wiley_logo.eps}}{}{}{\errmessage{arXiv: logo patch failed}}
\def\oddfoot@titlepage@info{%
  \vbox{\hsize\textwidth
    \noindent\rule{\textwidth}{.5\@p@t}\par\vskip2\@p@t
    \noindent{\fontsize{7}{8}\selectfont\hfill{\pagenumfont\thepage\ of \@LastPg}}%
    \vskip10\@p@t}}%
\def\evenfoot@titlepage@info{\oddfoot@titlepage@info}%
\def\evenhead@headings@info{\head@headings@def{\evenpage@headings@def\hfill}}%
\makeatother

\begin{document}
\raggedbottom

\title{Physics-Guided Crystal Structure Elucidation from Powder X-Ray Diffraction with Universal Machine-Learning Interatomic Potentials}
\author[1]{Yunsung Lim}
\author[2]{Seungwoo Hwang}
\author[1,2,3]{Seungwu Han}

\authormark{Lim \textsc{et al.}}
\titlemark{Physics-Guided Crystal Structure Elucidation from Powder X-Ray Diffraction}

\address[1]{\orgdiv{Research Institute of Advanced Materials}, \orgname{Seoul National University}, \orgaddress{\city{Seoul}, \postcode{08826}, \country{Republic of Korea}}}
\address[2]{\orgdiv{Department of Materials Science and Engineering}, \orgname{Seoul National University}, \orgaddress{\city{Seoul}, \postcode{08826}, \country{Republic of Korea}}}
\address[3]{\orgdiv{Center for AI and Natural Sciences}, \orgname{Korea Institute of Advanced Study}, \orgaddress{\city{Seoul}, \postcode{02455}, \country{Republic of Korea}}}

\corres{Seungwu Han (\email{hansw@snu.ac.kr})}


\abstract[ABSTRACT]{Determining unique atomic coordinates from powder X-ray diffraction (PXRD) remains challenging because peak overlap, weak scattering from light elements, limited scattering contrast between elements with similar atomic numbers, and experimental noise render the inverse problem underdetermined. Here, we present XCAPE (XRD-Constrained Accelerated Prediction Engine), a crystal structure prediction (CSP) framework that integrates experimentally derived composition, unit cell parameters, space group, and PXRD data into a symmetry-aware evolutionary search. XCAPE explores Wyckoff-site configurations using tailored mutation operations, evaluates candidates by both energetic stability and diffraction agreement, and accelerates geometry optimization using a universal machine-learning interatomic potential (MLIP). The framework recovered 71 of 77 database-derived benchmark structures (92.2\%) from simulated PXRD patterns and all 12 reference structures from experimental PXRD patterns. It further identified 12 candidate structure assignments that are consistent with experimental PXRD patterns, lack matching entries in the Inorganic Crystal Structure Database (ICSD), and are thermodynamically competitive at the MLIP level. Density functional theory (DFT) calculations showed that 10 of these 12 candidates lie within 10~meV/atom of the convex hull. By bridging experimental diffraction data and physically plausible atomistic models, XCAPE can provide structural insight into previously unresolved materials and support accelerated materials discovery in automated research workflows.}

\maketitle

\section{Introduction}

The discovery and development of advanced materials rely fundamentally on determining their atomic structures~\cite{structure_disc_1,csp}, as a material's properties are governed by the identities and spatial arrangements of its constituent atoms~\cite{structure_prop_1}. Accurate structural information therefore provides a basis for understanding structure--property relationships and designing improved materials through strategies such as elemental substitution~\cite{elem_substitution}, defect engineering~\cite{defect}, and phase stabilization~\cite{entropy_stabilized}. Among the experimental techniques available for crystal structure characterization, powder X-ray diffraction (PXRD) is particularly attractive because it is rapid, widely accessible, and applicable to polycrystalline samples~\cite{pxrd}. A PXRD pattern contains essential information about the unit cell, crystal symmetry, and atomic arrangement, making it a primary source of experimental constraints for crystal structure determination~\cite{pxrd_structure_1,pxrd_structure_2}.

Nevertheless, recovering a complete atomic structure from a PXRD pattern remains a challenging inverse problem. Since powder diffraction projects three-dimensional reciprocal-space information onto a one-dimensional pattern, distinct reflections frequently overlap~\cite{pxrd_hard_1}. Although unit cell parameters and plausible space groups can often be inferred from peak positions, determining atomic coordinates is substantially more difficult because they are encoded in diffraction intensities that are affected by peak overlap and experimental artifacts~\cite{pxrd_hard_2,pxrd_hard_3}. Consequently, different atomic arrangements can produce similar PXRD patterns, making diffraction agreement alone insufficient to unambiguously identify a structure~\cite{leeman_challenges}. For example, some entries in the Powder Diffraction File (PDF) maintained by the International Centre for Diffraction Data (ICDD)~\cite{icdd_5} report unit cell parameters and space groups but lack atomic coordinates. These limitations motivate the incorporation of complementary physical criteria to distinguish between structures consistent with the available diffraction data.

Energy-based crystal structure prediction (CSP) offers a promising route to addressing this ambiguity. Conventional CSP methods explore the potential energy surface (PES) to identify energetically favorable structures for a specified chemical composition~\cite{uspex,csp,spinner}. In PXRD-driven structure determination, experimental unit cell parameters and space group assignments can constrain this search, while calculated energies provide an additional criterion for assessing the physical plausibility of candidate atomic arrangements~\cite{xrd_prototype,fpass_1,fpass_2}. Combining diffraction agreement with energetic evaluation can therefore help discriminate between candidates that yield similar PXRD patterns. A high-throughput implementation of this strategy was provided by Griesemer et al.~\cite{xrd_prototype}, who developed a prototype-based method for assigning structures to ICDD entries lacking atomic coordinates. Candidate prototypes matching the target stoichiometry, experimental space group, and number of formula units per unit cell ($Z$) were retrieved from the Open Quantum Materials Database (OQMD)~\cite{oqmd}, decorated with the target elements, and placed on the experimental unit cell. The resulting candidates were evaluated using density functional theory (DFT) calculations and a residual factor (RF) measuring agreement with the experimental PXRD pattern, and the lowest-energy structure among those with RF values within 0.2 of the minimum was selected. Although effective in reducing the search space, this approach is restricted to structural prototypes represented in the reference database.

Database-independent approaches expand the range of structures accessible to PXRD-driven searches. Parackal et al.~\cite{wyckoff_prototype} adopted structural representations based on Wyckoff-site occupations. A \textit{Wyckoff prototype} specifies a space group and the occupied Wyckoff sites grouped by distinct but unspecified species, whereas a \textit{Wyckoff protostructure} (WPS) additionally assigns an element to each occupied site. The free positional parameters remain unspecified, allowing each WPS to represent a family of atomic structures. Parackal et al.~\cite{wyckoff_prototype} systematically enumerated composition- and symmetry-compatible WPSs, ranked them using formation energies predicted by a pretrained machine-learning model, and optimized candidate structures through gradient-based fitting to experimental PXRD patterns before final validation using DFT calculations. In a complementary approach, Lee et al.~\cite{evolvmorph} developed \textit{Evolv\&Morph}, which combines an evolutionary algorithm with Bayesian optimization to generate structures that maximize the similarity between simulated and target diffraction patterns. These studies demonstrate how symmetry-based representations and global optimization can extend structure determination beyond existing structural databases.

Within a symmetry-constrained search, identifying a candidate structure involves both selecting the Wyckoff sites occupied by each element and determining the free positional parameters associated with those sites. Experimental symmetry constraints reduce the search space, but multiple composition-compatible WPSs and their associated atomic configurations may still need to be explored. This motivates search strategies that can move efficiently between different Wyckoff-site occupations while optimizing the atomic arrangements represented by each WPS. Incorporating energy-based evaluation into this search introduces an additional computational cost, particularly when repeated geometry optimizations are performed using DFT. Efficient exploration of WPSs and rapid energetic evaluation are therefore complementary requirements for making energy-based, PXRD-constrained structure searches practical.

Here, we develop \textit{XCAPE} (XRD-Constrained Accelerated Prediction Engine), a PXRD-driven, symmetry-constrained CSP framework for identifying energetically plausible atomic structures consistent with experimental diffraction data. XCAPE combines structure generation constrained by the experimentally determined unit cell and space group with an evolutionary programming algorithm for global optimization. We devise mutation operations that act on WPSs, enabling exploration of alternative Wyckoff-site occupations compatible with the target composition and symmetry. To accelerate energetic evaluation and the repeated geometry optimizations required during the search, we employ SevenNet-Omni~\cite{7net_omni,sevennet}, a universal machine-learning interatomic potential (MLIP), in place of computationally demanding DFT calculations~\cite{mlip_1,mlip_2,mlip_csp,spinner}. Together, WPS-based mutations and MLIP-driven optimization enable efficient exploration of candidate atomic arrangements under experimental structural constraints. We validate XCAPE using both PXRD patterns simulated from known structures in the Materials Project (MP)~\cite{mp_db} and the Inorganic Crystal Structure Database (ICSD)~\cite{icsd} and experimental PXRD patterns from the ICDD database~\cite{icdd_5}.

\section{Results}

\subsection{Overview of the XCAPE framework}\label{sec:overview}
\begin{figure*}
\centering
  \includegraphics[width=\textwidth]{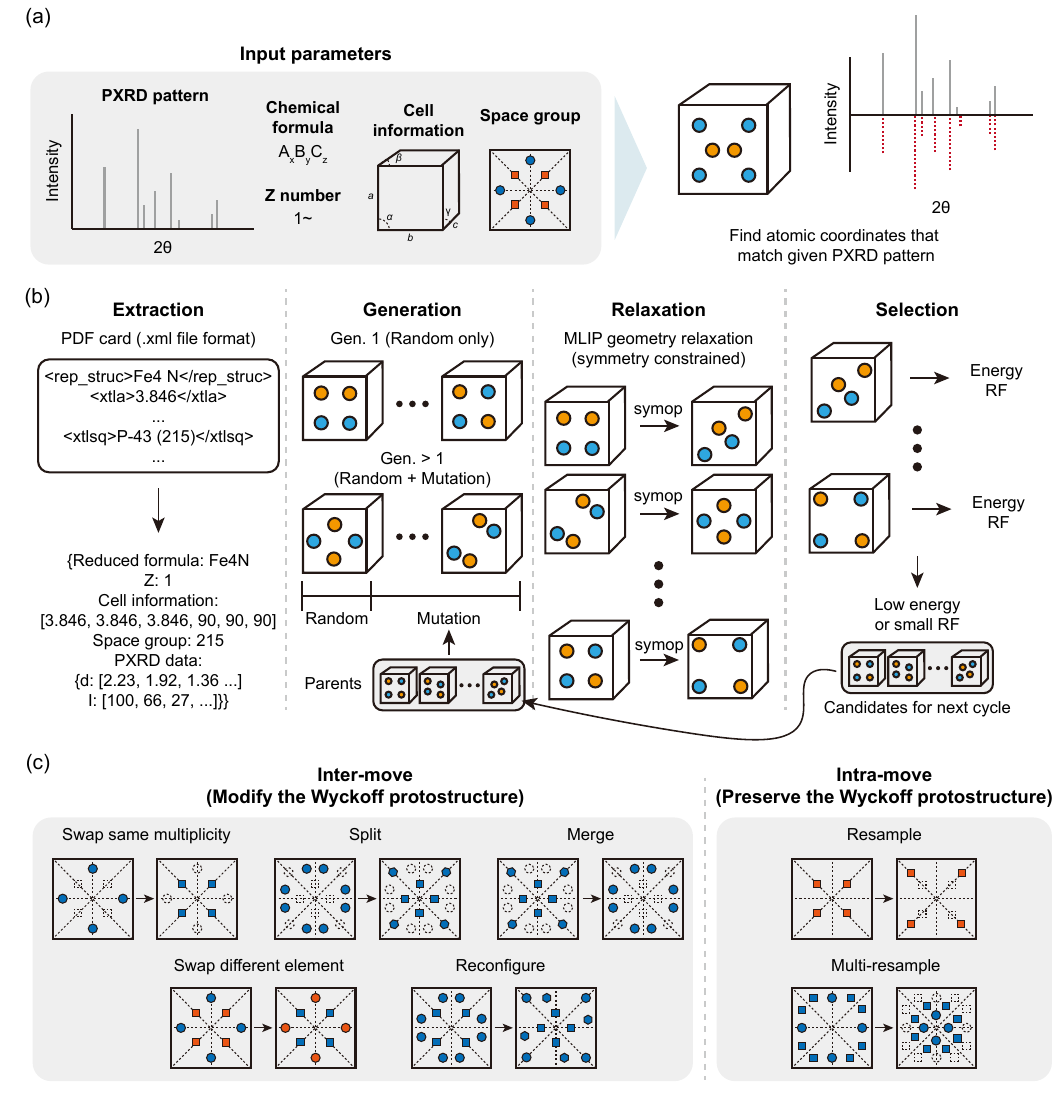}
  \caption{
  \textbf{Schematic overview of XCAPE.}
  \textbf{(a)} Objective of XCAPE. The PXRD pattern, chemical composition, number of formula units per unit cell ($Z$), unit cell parameters, and space group are provided as inputs, and XCAPE predicts atomic coordinates consistent with these constraints.
  \textbf{(b)} XCAPE workflow comprising four stages: (1) input extraction, (2) structure generation, (3) geometry relaxation, and (4) candidate selection.
  \textbf{(c)} Symmetry-aware mutation operations used in the evolutionary search. Five inter-moves modify the occupied Wyckoff-site pattern and/or elemental assignments, whereas two intra-moves preserve the Wyckoff protostructure while resampling its free positional parameters. Colors indicate chemical species, and marker shapes indicate Wyckoff sites.
  }
  \label{fgr:xcape}
\end{figure*}

XCAPE is a Python package designed to predict atomic arrangements consistent with experimental PXRD data and known crystallographic information. As illustrated in Figure~\ref{fgr:xcape}a, XCAPE takes a PXRD pattern, chemical formula, $Z$, unit cell parameters, and space group as inputs. By treating the composition, $Z$, unit cell, and space group as fixed constraints, the structure determination problem is reduced to a targeted search over the unknown atomic coordinates. In particular, the space group commonly reported in PDF entries restricts the configurational search space.

Figure~\ref{fgr:xcape}b summarizes the evolutionary programming workflow of XCAPE, which comprises four stages: (1) input extraction, (2) structure generation, (3) geometry relaxation, and (4) candidate selection. First, the crystallographic information and PXRD data are extracted from a PDF entry. XCAPE then generates symmetry-compatible structures using the PyXtal~\cite{pyxtal} Python package. All structures in the initial generation are generated randomly, whereas subsequent generations contain both randomly generated structures and offspring obtained by mutating selected parent structures. Each generated structure undergoes symmetry-constrained geometry optimization using a universal MLIP. During both structure generation and relaxation, structures that violate the minimum allowed interatomic distance for any pair of elements are discarded (see Note~S1). The relaxed structures are evaluated according to their predicted energies and RF values. Candidates with either low energies or small RF values are retained as parents for the next generation, allowing both objectives to guide the subsequent search.

To efficiently explore the symmetry-constrained configurational space, XCAPE employs seven mutation operations divided into two classes (see Figure~\ref{fgr:xcape}c). \textit{Inter-moves} modify the WPS by changing the occupied Wyckoff sites or their elemental assignments, and \textit{intra-moves} preserve the WPS but alter the symmetry-allowed positional parameters.

We devised five inter-moves, as illustrated in the left panel of Figure~\ref{fgr:xcape}c. A \textit{swap same multiplicity} replaces an occupied Wyckoff site with another site of the same multiplicity, for example, Li(4a) \(\rightarrow\) Li(4b). A \textit{split} replaces one Wyckoff site with two sites whose multiplicities sum to that of the original site, such as Li(8c) \(\rightarrow\) Li(4a) + Li(4b), whereas a \textit{merge} performs the reverse operation. A \textit{swap different element} exchanges the elemental assignments of two occupied Wyckoff sites with the same multiplicity, for example, Li(4a) + Mg(4b) \(\rightarrow\) Mg(4a) + Li(4b). Finally, a \textit{reconfigure} replaces the set of Wyckoff sites occupied by a given element with another multiplicity-preserving combination, such as Li(2i) + Li(1a) \(\rightarrow\) Li(1b) + Li(1c) + Li(1d). This operation was introduced to connect regions of the discrete WPS space that cannot be directly linked by the other four inter-moves.

We evaluated the reachability of the WPS space using a breadth-first search on 21 structures from the MP database (see Note~S2 and Table~S1). For each test case, feasible WPSs were represented as graph nodes, and pairs of WPSs connected by an allowed inter-move were represented as edges. As shown in Figure~S1, the coverage reached 100\% for every test case. Thus, within the WPS spaces examined here, any enumerated WPS could be reached from any other through a finite sequence of the five inter-moves.

The two intra-moves, shown in the right panel of Figure~\ref{fgr:xcape}c, explore the continuous degrees of freedom within a fixed WPS. \textit{Resample} modifies the free positional parameters of a single occupied Wyckoff site, whereas \textit{multi-resample} simultaneously modifies those of multiple occupied sites to induce a larger structural perturbation. Together, the seven operations enable XCAPE to traverse both the discrete space of WPSs and the continuous space of symmetry-allowed atomic coordinates.

\subsection{Benchmarking against known crystal structures}\label{sec:benchmark}

Before applying XCAPE to experimental PXRD data, we evaluated its structure-recovery capability using reference materials for which the atomic structures were known. For each material, the target PXRD pattern was calculated from the reference structure, and its chemical formula, $Z$, unit cell parameters, and space group were provided to XCAPE, while the atomic coordinates were withheld. The benchmark set comprised 77 materials: 50 ternary oxides, nitrides, halides, and sulfides sampled from the MP database~\cite{mp_db} (see Table~S1 for the sampling details), 21 ternary structures from the ICSD~\cite{icsd} reported by Kang et al.~\cite{spinner}, and six quaternary solid electrolytes from the ICSD reported by Hwang et al.~\cite{topic}. The ICSD structures were added to include more challenging systems. The structures from Kang et al.\ include low-symmetry cases and six for which the exact reference structures were not recovered in the previous study. The solid electrolytes extend the benchmark to compositions with more elements and unit cells containing over 100 atoms, in which the positions of mobile Li ions can be difficult to resolve~\cite{pxrd,cussen_garnet}.

As summarized in Figure~S2, the resulting benchmark spans broad ranges of structural complexity, including the number of atoms per conventional unit cell, the space group, and the number of possible WPSs. Because both the target PXRD patterns and their underlying reference structures were available, this benchmark allowed us to quantitatively assess whether XCAPE could recover the known atomic arrangements under controlled, noise-free conditions.

\begin{figure*}
\centering
  \includegraphics[width=\textwidth]{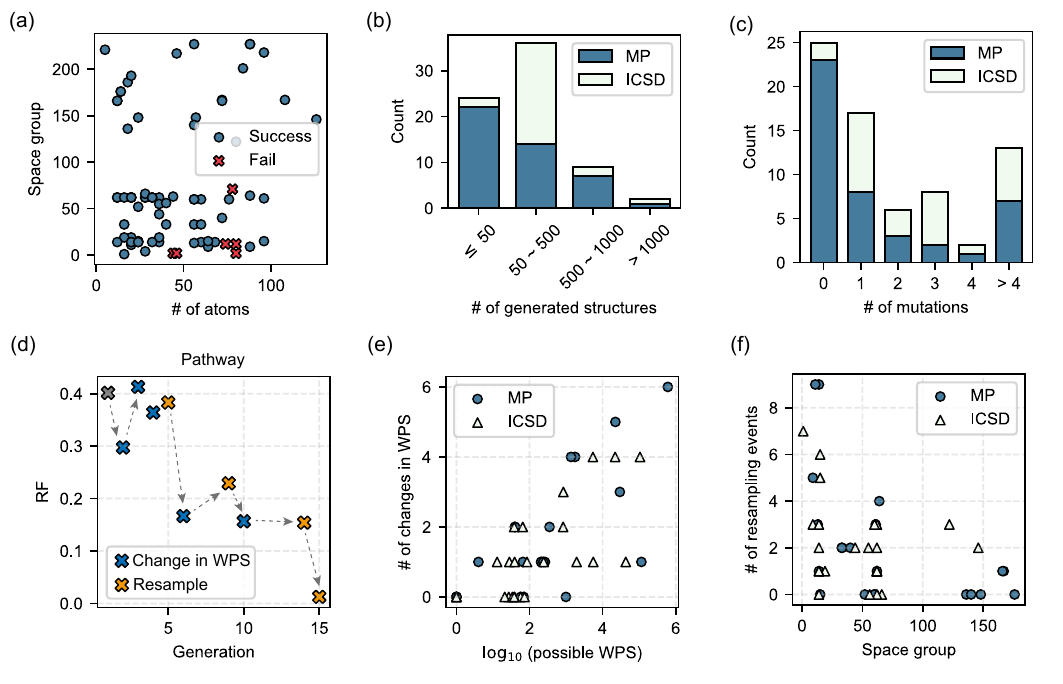}
\caption{
  \textbf{Performance of XCAPE on the MP and ICSD benchmark sets.}
  \textbf{(a)} Success and failure outcomes as functions of the number of atoms in the conventional cell and the space group number.
  \textbf{(b)} Distribution of the number of generated candidate structures required to recover the reference structure.
  \textbf{(c)} Distribution of the number of mutation operations applied before recovery of the reference structure.
  \textbf{(d)} Representative search trajectory leading to the reference structure. Gray markers indicate the initial candidate, blue markers indicate candidates obtained after changes in the Wyckoff protostructure, and orange markers indicate candidates obtained through positional resampling without changing the Wyckoff protostructure.
  \textbf{(e)} Relationship between the number of possible Wyckoff protostructures and the number of Wyckoff protostructure changes required to recover the reference structure.
  \textbf{(f)} Relationship between the space group number and the number of positional resampling operations that preserve the Wyckoff protostructure.
}
  \label{fgr:benchmark}
\end{figure*}

Among the 77 benchmark cases, XCAPE successfully recovered 71 reference structures within 100 generations, corresponding to an overall success rate of 92.2\%. As shown in Figure~\ref{fgr:benchmark}a, five of the six failed cases were low-symmetry structures with space group numbers of 12 or lower, and all six contained more than 40 atoms in their conventional unit cells. Reduced symmetry generally increases the number of crystallographically independent sites, while a larger unit cell increases the number of atoms that must be assigned to those sites. The combination of these factors substantially enlarges both the discrete WPS space and the continuous coordinate space, suggesting that the search budget of 100 generations (50 structures per generation) was insufficient for these particularly complex cases.

Most structures, however, were recovered well within the 100-generation budget (see Figure~\ref{fgr:benchmark}b, c, and Table~S1). Of the complete benchmark set, 24 cases (31.2\%) were solved directly from the randomly generated initial population, without requiring any mutation, and 60 cases (77.9\%) were solved within the first 10 generations. With the population size used here, this corresponds to the evaluation of no more than 500 candidate structures. More than half of the successful cases required no more than one mutation event along the successful lineage. However, 15 cases (21.1\% of the successful cases) required four or more mutations, indicating that iterative evolutionary exploration remained essential for a non-negligible subset of the benchmark.

To examine how mutation operations contributed to structure recovery, we traced the successful lineage of each predicted structure and recorded whether each mutation produced an actual change in its WPS (see Figure~\ref{fgr:benchmark}d as an example). This distinction is necessary because a stochastic inter-move can occasionally regenerate the same Wyckoff-site combination while changing only its positional parameters. As shown in Figure~\ref{fgr:benchmark}e, the number of WPS-changing events generally increased with the number of possible WPSs. This trend indicates that systems with larger discrete WPS spaces require more extensive exploration of alternative Wyckoff-site combinations. In contrast, positional resampling events were particularly frequent for low-symmetry structures (see Figure~\ref{fgr:benchmark}f). Since such structures have more independent atomic coordinates, repeated sampling is required to generate a starting configuration that relaxes to the reference structure upon MLIP-based geometry optimization. These results show that the inter- and intra-moves address complementary sources of complexity: the former explores discrete Wyckoff-site assignments, whereas the latter explores continuous symmetry-allowed atomic coordinates.

Because XCAPE records the complete mutation history of each search, we further analyzed how individual mutation operations contributed to successful structure recovery (see Figure~S3). The reconfigure, resample, and multi-resample operations collectively accounted for 60\% of the mutation events along the successful trajectories, because they remain applicable even when few alternative WPSs are available. For trajectories containing three or more mutation events, inter-moves dominated the early stage of the search, whereas intra-moves became more frequent during the middle stage. In the late stage, reconfigure was the most frequent individual operation. This sequence suggests that XCAPE first explores discrete WPSs broadly, then refines atomic coordinates within promising protostructures, and uses reconfigure to escape unproductive search basins.

\subsection{Robustness and search efficiency of XCAPE}\label{sec:robustness}

\begin{figure*}
\centering
  \includegraphics[width=\textwidth]{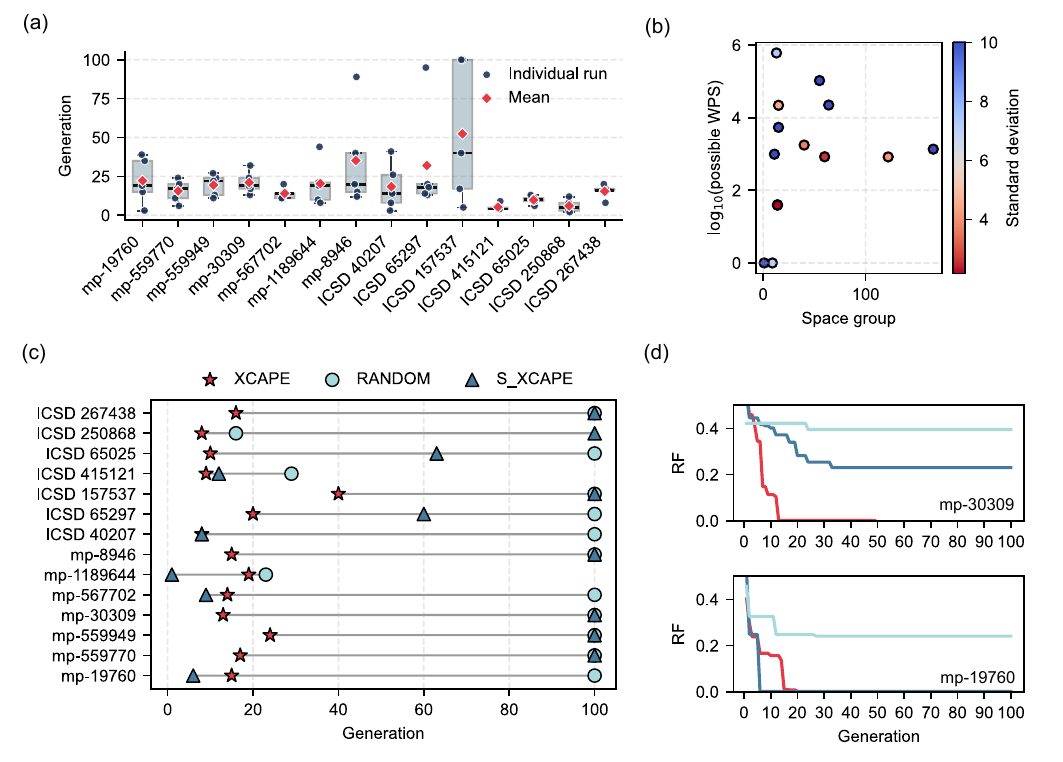}
\caption{
  \textbf{Robustness and comparative performance of XCAPE.}
  \textbf{(a)} Run-to-run variability of XCAPE for the selected benchmark cases. Navy circles represent the results of five independent runs, red diamonds indicate their mean values, and box plots summarize their distributions.
  \textbf{(b)} Dependence of the run-to-run variability on the space group number and the number of possible Wyckoff protostructures. Colors indicate the standard deviation of the recovery generation across the five runs. Values above 10 are shown in the darkest color.
  \textbf{(c)} Comparison of XCAPE with random-only search and S\_XCAPE. Red stars, lightblue circles, and blue triangles represent XCAPE, random-only search, and S\_XCAPE, respectively. Performance is reported as the number of generations required to recover the reference structure in a single run of each method. For XCAPE, the original benchmark run is shown. In (a) and (c), a value of 100 indicates that the reference structure was not recovered within the search limit.
  \textbf{(d)} Representative RF trajectories for mp-30309 and mp-19760, for which XCAPE and S\_XCAPE, respectively, recovered the reference structure considerably faster than the other methods. Line colors follow the legend in (c).
}
  \label{fgr:performance}
\end{figure*}

Before comparing XCAPE with alternative search strategies, we assessed its run-to-run robustness using the benchmark cases that required eight or more generations in the original run. Of the 15 such cases, mp-18341 was excluded because its reference structure was recovered from a randomly generated candidate rather than through mutation, leaving 14 cases. Four additional independent runs were performed for each case, and searches that did not recover the reference structure were recorded at generation 100 (see Figure~\ref{fgr:performance}a).

For most cases, the generation at which the reference structure was recovered remained reasonably consistent across independent runs, with ICSD 157537~\cite{icsd_157537} showing the largest variation. The standard deviation in the recovery generation ranged from 2.17 to 45.24 generations, with a mean of 13.96, and 11 of the 14 cases exhibited standard deviations below 20 generations. Within this subset of cases recovered in the original benchmark run, the run-to-run variation showed no clear dependence on the number of possible WPSs (see Figure~\ref{fgr:performance}b). Thus, although low-symmetry structures with large unit cells were more difficult to recover, a larger discrete WPS space did not noticeably increase the stochastic variability of the search for cases that XCAPE could recover.

We next compared XCAPE with a random-only search and a search strategy adopted from SPINNER~\cite{spinner}, hereafter denoted S\_XCAPE. In S\_XCAPE, the XCAPE mutation operations were replaced by the permutation operator of SPINNER, and geometry optimization was conducted without symmetry constraints because permutation does not preserve the assigned space group symmetry (see Methods). As shown in Figure~\ref{fgr:performance}c, XCAPE recovered the reference structure more efficiently than the random-only search and S\_XCAPE in 10 of the 14 cases, and XCAPE and S\_XCAPE performed equally for ICSD 40207~\cite{icsd_40207}. The random-only search generally failed to recover the reference structures within the search budget, demonstrating the importance of evolutionary inheritance and mutation (see Figure~S4). S\_XCAPE nevertheless showed substantially improved performance for three cases (mp-19760, mp-567702, and mp-1189644; see Figures~\ref{fgr:performance}d and S5). When applied to the six cases not solved by the original XCAPE workflow, S\_XCAPE recovered only mp-18549, which has a relatively high space group number (see Figure~S6).

To identify the origin of this improvement, we traced the successful pathways of the four cases for which S\_XCAPE performed better. In all four cases, the recovered structures originated from randomly generated candidates rather than permutation-derived offspring. This observation suggests that unconstrained geometry optimization, rather than the permutation operator itself, contributed to the improved recovery, although the larger fraction of randomly generated candidates in S\_XCAPE may also have played a role. Symmetry-constrained optimization efficiently refines candidates already located within an appropriate basin, but it may prevent access to relaxation pathways involving transient symmetry breaking that could otherwise connect the initial candidate to the basin of the reference structure. We therefore introduced an optional hybrid relaxation scheme in which a prescribed fraction of the low-energy candidates in each generation are additionally relaxed without symmetry constraints (see Methods). We tested this scheme on three of the four cases, excluding mp-1189644, which was also solved by the random-only search. All three reference structures were recovered in fewer generations than with the original XCAPE settings (see Figure~S7). Accordingly, this hybrid relaxation scheme was used in all subsequent XCAPE calculations unless otherwise specified.

\subsection{Validation against experimental PXRD patterns}\label{sec:valid_exp}

\begin{figure*}
\centering
  \includegraphics[width=\textwidth]{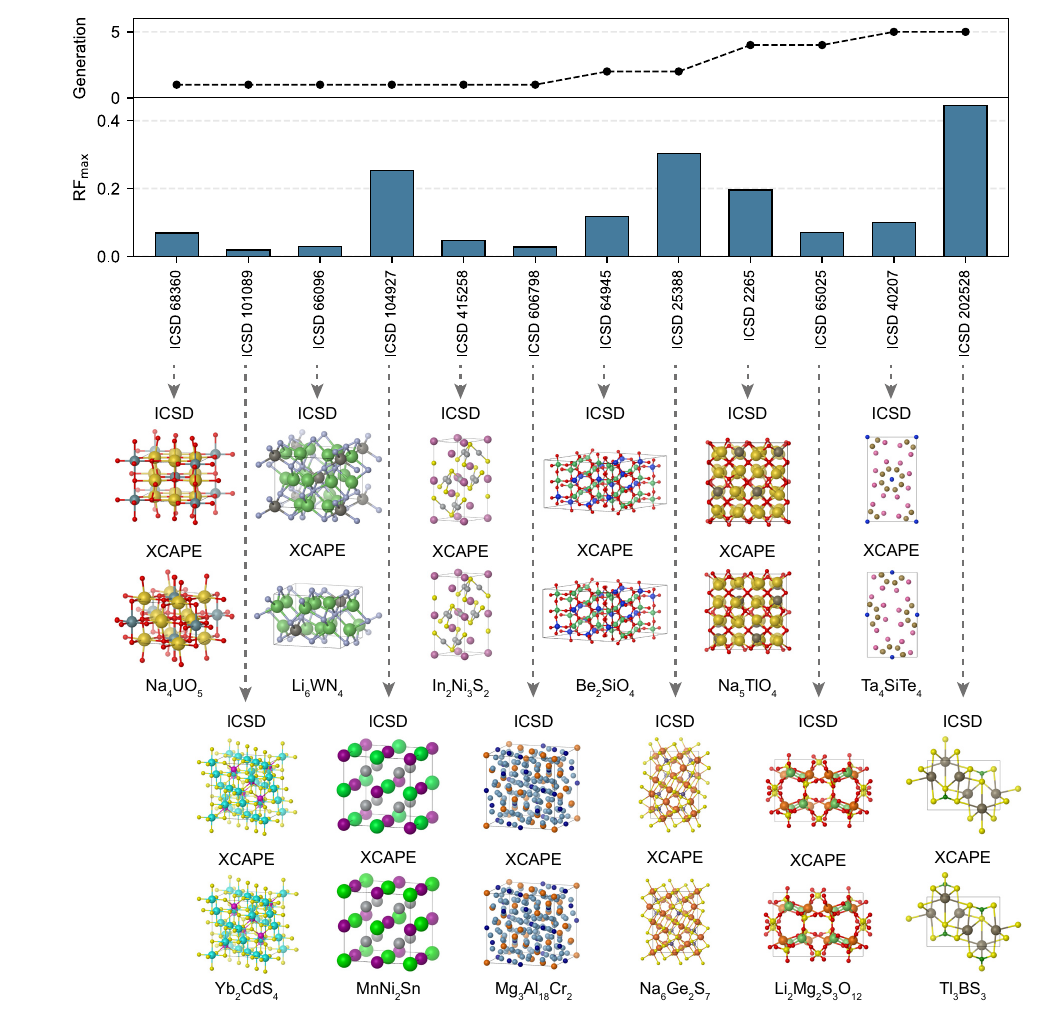}
  \caption{
  \textbf{Practical validation of XCAPE using experimental PXRD patterns.}
  For each of the 12 validation cases, the number of generations required to recover each reference structure is reported together with the maximum RF observed among the successfully recovered structures from XCAPE. The experimental reference structure obtained from the ICSD and the corresponding structure recovered by XCAPE are shown for comparison.
}
  \label{fgr:benchmark_exp}
\end{figure*}

We next evaluated XCAPE using experimental PXRD patterns for 12 cases in which both a PDF entry and a corresponding crystal structure in the ICSD were available. Two of these cases were also included in the preceding simulated-pattern benchmark. Detailed information on the 12 cases is provided in Table~S2. XCAPE was run for up to 100 generations for each case, and the corresponding ICSD reference structure was successfully recovered in all 12 cases. The results for all 12 cases are shown in Figure~\ref{fgr:benchmark_exp}.

Unlike the preceding benchmark, in which the target PXRD patterns were calculated directly from the reference structures, the experimental patterns did not perfectly match the patterns calculated from the corresponding ICSD structures. The RF values of the recovered reference structures ranged from 0.02 to 0.44. Such discrepancies can arise from experimental noise, preferred orientation, sample-preparation conditions, and minor impurities~\cite{pxrd}, as well as differences between the measured sample and its reported structural model. For example, for ICSD 104927~\cite{icsd_104927}, all converged candidates generated during the XCAPE search were structurally equivalent to the ICSD reference structure, yet their RF values remained 0.25 (see Figure~S8). This residual reflects a mismatch between the PDF diffraction data and the calculated pattern, rather than a failure to recover the structure.

To assess the range of RF values associated with correct structures, we extended this comparison to 100 ICSD--PDF pairs, including the 12 validation cases (see Note~S3, Figure~S9, and Table~S3). Since XCAPE constrains candidates to the PDF unit cell, each ICSD structure was mapped onto the corresponding PDF lattice before calculating RF. All 100 reference structures yielded RF values of at most 0.44. We therefore adopted an RF threshold of 0.45 to retain correct structures despite experimental discrepancies. This threshold was used only for candidate selection in the subsequent analysis, whereas recovery in the 12 validation cases was determined by direct structural comparison with the ICSD reference structures.

Importantly, the candidate with the smallest RF was not necessarily the most physically plausible structure. As shown in Figure~S8, several candidates exhibited RF values lower than that of the recovered reference structure but had substantially higher predicted energies. Selection based solely on diffraction agreement can therefore favor energetically unfavorable structures that overfit features of the experimental pattern. To balance diffraction agreement and energetic plausibility, XCAPE reports two complementary solutions. The stability-prioritized solution is defined as the lowest-energy candidate among those with RF $\leq$ 0.45. Conversely, the diffraction-prioritized solution is defined as the candidate with the smallest RF among those satisfying $E-E_{\mathrm{min}}\leq 50$~meV/atom, where $E_{\mathrm{min}}$ is the lowest predicted energy among all candidates generated during the search. This energy window was chosen to lie within the range typical of experimentally observed metastable inorganic compounds, whose energies above the convex hull have a median of 15~meV/atom and a 90th percentile of 67~meV/atom~\cite{sun_metastability}. Together, these criteria retain candidates that are consistent with the experimental diffraction pattern while excluding solutions that are strongly disfavored on the predicted PES. In the following section, we apply these two selection criteria to revisit the cases previously examined by Griesemer et al.~\cite{xrd_prototype}.

\subsection{Revisiting structure assignments from experimental PXRD patterns}\label{sec:revisit}

\begin{figure}
\centering
  \includegraphics[width=\columnwidth]{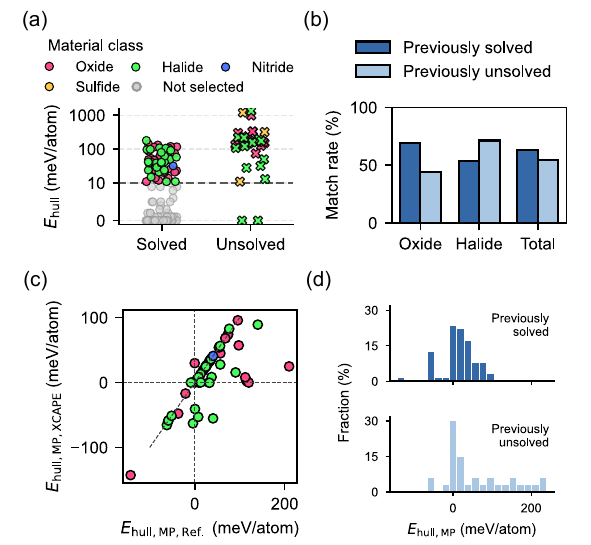}
  \caption{
  \textbf{Selection of target cases from the previous study and comparison with XCAPE results.}
  \textbf{(a)} Distributions of $E_{\mathrm{hull}}$ reported in the previous study for cases classified as solved (circles) or unsolved (crosses). Gray markers denote solved cases that were not selected, and the dashed line indicates the $E_{\mathrm{hull}}$ threshold (10~meV/atom) used to select cases for re-evaluation with XCAPE.
  \textbf{(b)} Agreement rates between the XCAPE candidates and the structural assignments reported in the previous study. Because the nitride and sulfide subsets contained only one and three cases, respectively, only the oxide, halide, and overall results are shown.
  \textbf{(c)} Comparison of $E_{\mathrm{hull, MP}}$ between the previously reported reference structures and the corresponding XCAPE candidates. Reference structures are labeled ``Ref.'' in the plot. 
  \textbf{(d)} Distributions of $E_{\mathrm{hull, MP}}$ for XCAPE candidates grouped according to whether the corresponding cases were classified as solved or unsolved in the previous study.
}
  \label{fgr:revisit}
\end{figure}

To further assess the practical utility of XCAPE, we revisited the dataset examined by Griesemer et al.~\cite{xrd_prototype} and investigated whether XCAPE could recover the reported structures or identify alternative candidate solutions. That study reported new structure assignments for 521 of the 624 candidate structures considered. Consistent with the chemical scope of our preceding benchmarks, we restricted the present analysis to ternary oxides, nitrides, halides, and sulfides composed of two metallic elements and a single anion species. Within this subset, we selected all 33 cases that the previous study did not deem solved because their prototype solutions failed its energy and/or RF validation checks, together with 63 solved cases whose reported structures had energies above the convex hull ($E_{\mathrm{hull}}$) greater than 10~meV/atom. These solved cases were included because their relatively high hull energies suggested that more stable structures consistent with the same PXRD patterns might exist. In total, XCAPE was applied to 96 cases (see Figure~\ref{fgr:revisit}a). For brevity, PDF entry numbers are given without the common prefix ``00-'' (e.g., 026-0326 for PDF No.~00-026-0326).

The resulting dataset was strongly compositionally imbalanced, comprising 52 oxides, 40 halides, and only four nitrides or sulfides. For each case, we retained the stability-prioritized and diffraction-prioritized candidates according to the selection criteria defined in the preceding section. For previously solved cases, these candidates were compared with the reported complete structure. For unsolved cases, which lacked a complete reference structure, the candidates were instead compared manually with the prototype structures proposed in the previous study. A case was classified as matched if either of the two XCAPE-selected candidates agreed with the previously reported structure or prototype. This comparison was based on the XCAPE outputs before the full geometry relaxation used for the $E_{\mathrm{hull}}$ calculations. Using this criterion, as shown in Figure~\ref{fgr:revisit}b, XCAPE reproduced the earlier structure assignments for 40 of the 63 previously solved cases (63.5\%). Among the 33 previously unsolved cases, 18 (54.5\%) yielded at least one candidate consistent with the available prototype information. In contrast, the halide subset showed a higher agreement rate for the previously unsolved cases than for the solved cases. However, the aggregate result was dominated by the larger oxide subset.

For the previously solved cases, the reported structures enabled a direct energetic comparison with the XCAPE predictions. Both the previously reported structures and the XCAPE candidates were fully relaxed, including the lattice parameters and atomic coordinates, using the same protocol (see Methods). Hereafter, $E_{\mathrm{hull}}$ is evaluated using competing phases obtained from the MP database and is denoted $E_{\mathrm{hull, MP}}$. Of the two candidates returned by XCAPE, the one with the lower $E_{\mathrm{hull, MP}}$ was retained for this comparison. As shown in Figure~\ref{fgr:revisit}c, XCAPE generally identified structures with $E_{\mathrm{hull, MP}}$ values lower than or comparable to those reported in the previous study. Evaluated with the same RF protocol, the XCAPE candidates also exhibited RF values lower than or comparable to those of the previously reported structures (see Figure~S10). The XCAPE candidates for the previously unsolved cases exhibited a broad high-energy tail extending beyond 200~meV/atom, whereas those for the previously solved cases remained below approximately 100~meV/atom (see Figure~\ref{fgr:revisit}d). These high $E_{\mathrm{hull, MP}}$ values for the unsolved cases may reflect intrinsic metastability, competition with alternative phases, or the possibility that the correct low-energy structures have not yet been identified.

\begin{figure*}
\centering
  \includegraphics[width=\textwidth]{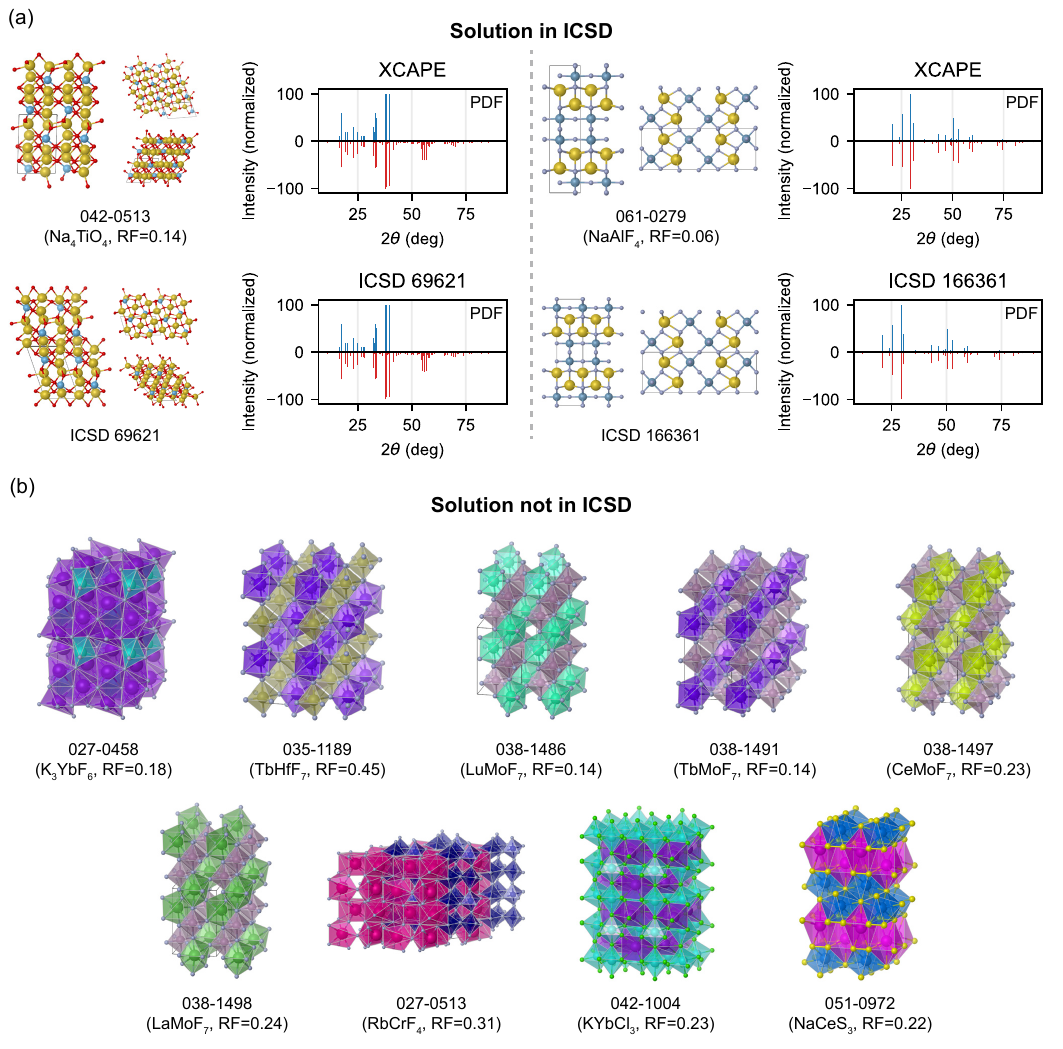}
\caption{
  \textbf{Crystal structure elucidation from experimental PXRD patterns using XCAPE.}
  \textbf{(a)} Two representative cases, PDF Nos.~042-0513 and 061-0279, for which corresponding reference structures are available in the ICSD. The XCAPE-recovered and ICSD reference structures are shown together with their PXRD patterns. In the right panels, blue and red lines denote the experimental patterns obtained from the PDF entries and the calculated patterns, respectively.
  \textbf{(b)} Nine candidate structures derived from experimental PXRD patterns for which no matching reference structures were identified in the ICSD. Each predicted structure is presented with its PDF number, reduced formula, and RF value.
}
  \label{fgr:revisit_2}
\end{figure*}

Among the 38 cases for which neither of the two XCAPE candidates matched the previous assignments, independently reported crystal structures were available in the ICSD for 15 cases. We identified these entries by searching the reduced formulas and/or chemical systems, and calculated their PXRD patterns using the same protocol employed in XCAPE. Comparison of the crystal structures and of the calculated and experimental PXRD patterns showed that XCAPE correctly recovered the ICSD structures for 14 of the 15 cases, none of which corresponded to the structures assigned in the previous study. In the remaining case (PDF No.~032-1069), the XCAPE candidate closely resembled the ICSD structure but exhibited minor structural differences. Neither structure fully reproduced the experimental pattern, although the XCAPE candidate showed a lower RF than the ICSD structure placed on the PDF unit cell (0.44 vs.\ 0.54). Two representative examples are shown in Figure~\ref{fgr:revisit_2}a, together with the XCAPE and ICSD structures and their corresponding PXRD patterns. The other matched cases are presented in Figures~S11 and S12.

For the remaining 23 of the 38 unmatched cases, no corresponding ICSD structures were identified. Applying the criteria RF $\leq$ 0.45 and $E_{\mathrm{hull, MP}}\leq 10$~meV/atom, together with an $E_{\mathrm{hull, MP}}$ lower than that of the previously reported structure for solved cases, we identified nine energetically plausible candidate structures that were consistent with the experimental PXRD patterns from the PDF entries. These candidates, illustrated in Figure~\ref{fgr:revisit_2}b, comprised eight halides and one sulfide. One of them (PDF No.~035-1189) lies at the RF threshold (RF $=$ 0.450). Notably, four were assigned to an \ch{AMoF7}-type prototype, where A denotes a lanthanide element. Detailed comparisons between their calculated and experimental PXRD patterns are provided in Figure~S13.

\subsection{Cross-fidelity evaluation and DFT validation}

The XCAPE benchmarks described above were performed using the \textsf{mpa} channel of SevenNet-Omni, which targets the MP-compatible Perdew--Burke--Ernzerhof (PBE) generalized gradient approximation (GGA) functional~\cite{pbegga} with Hubbard $U$ corrections~\cite{dudarev,jain_gga_u}, hereafter denoted PBE(+U). Because PBE tends to overestimate the equilibrium lattice parameters of many solids~\cite{pbe_over}, constraining candidates to the experimental unit cells from PDF entries could affect the MLIP-predicted energies and forces. In addition, a previous CSP study~\cite{spinner} reported that PBE can incorrectly stabilize predicted structures relative to experimentally observed ones and that recalculation with the meta-GGA SCAN functional~\cite{scan} corrected the energy ordering in four of the six such cases. We therefore recalculated $E_{\mathrm{hull, MP}}$ using the \textsf{matpes\_r2scan} channel of SevenNet-Omni~\cite{7net_omni}, trained on r$^{2}$SCAN~\cite{r2scan} calculations from the MatPES dataset~\cite{matpes}. Compared with PBE, r$^{2}$SCAN, a revised form of SCAN, predicts equilibrium volumes of solids in closer agreement with experiment~\cite{kothakonda_r2scan}. Both the XCAPE candidates and all competing phases were fully relaxed at the same fidelity (see Methods).

\begin{figure*}
\centering
  \includegraphics[width=\textwidth]{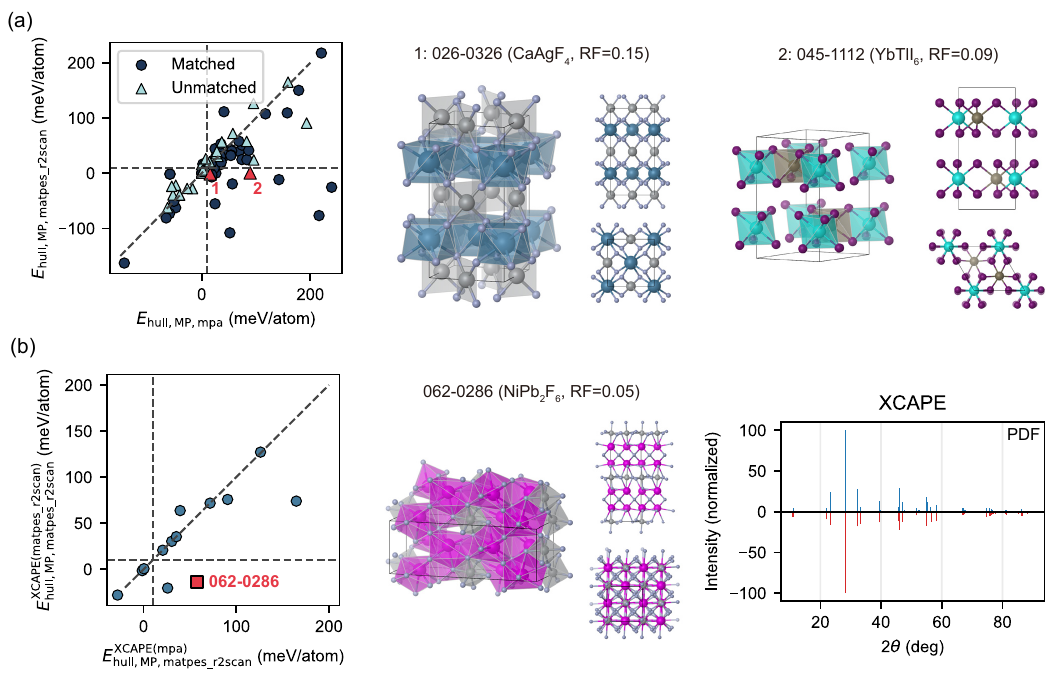}
  \caption{
  \textbf{Cross-fidelity evaluation of XCAPE candidates.}
  \textbf{(a)} Comparison of PBE(+U)-fidelity (\textsf{mpa}) and r$^{2}$SCAN-fidelity (\textsf{matpes\_r2scan}) $E_{\mathrm{hull, MP}}$ values for candidates generated by PBE(+U)-fidelity XCAPE searches. Circles and triangles denote cases in which the XCAPE candidates matched and did not match the structure assignments reported in the previous study, respectively. Red triangles highlight unmatched cases that fall within 10~meV/atom of the convex hull only at the r$^{2}$SCAN fidelity. Dashed lines indicate $E_{\mathrm{hull, MP}} = 10$~meV/atom and parity. The corresponding crystal structures are shown in the middle and right panels together with their reduced formulas and RF values.
  \textbf{(b)} Comparison of r$^{2}$SCAN-fidelity $E_{\mathrm{hull, MP}}$ values for candidates obtained from PBE(+U)-fidelity (horizontal axis) and r$^{2}$SCAN-fidelity (vertical axis) XCAPE searches. The red square highlights a promising candidate recovered only by the r$^{2}$SCAN-fidelity search. Its predicted crystal structure is shown in the middle panel, and the experimental PXRD pattern from the PDF entry (blue) is compared with the calculated pattern of the candidate (red) in the right panel.
}
  \label{fgr:r2scan}
\end{figure*}

As shown in the left panel of Figure~\ref{fgr:r2scan}a, the r$^{2}$SCAN-fidelity calculations yielded lower $E_{\mathrm{hull, MP}}$ values for many structures that appeared metastable at the PBE(+U) fidelity. For this comparison, the lower $E_{\mathrm{hull, MP}}$ of the two XCAPE candidates was taken for each case at each fidelity. In particular, 57 of the 96 cases had $E_{\mathrm{hull, MP}}>10$~meV/atom at the PBE(+U) fidelity, and 14 of these shifted into the near-hull region (defined here as $E_{\mathrm{hull, MP}}\leq10$~meV/atom). Two of these candidates, PDF Nos.~026-0326 and 045-1112, did not match the structure assignments reported in the previous study, and no corresponding structures were identified in the ICSD. Their agreement with the experimental PXRD patterns (see Figure~S13) and low r$^{2}$SCAN-fidelity hull energies therefore support them as promising alternative structure assignments. The proposed structures for these two cases are illustrated in the middle and right panels of Figure~\ref{fgr:r2scan}a.

Since MLIP fidelity can affect not only the predicted stability of a candidate but also the search trajectory, we repeated XCAPE using the r$^{2}$SCAN-fidelity model for selected cases. These cases met the following three conditions: (1) the XCAPE candidate did not match the structural assignment reported in the previous study, (2) no corresponding structure was identified in the ICSD, and (3) the PBE(+U)-fidelity $E_{\mathrm{hull, MP}}$ exceeded 10~meV/atom. Of the 12 cases meeting these criteria, PDF No.~042-0410 was excluded because its candidate matched the previously reported structure after geometry relaxation. Two cases omitted from the preceding analysis were added: PDF No.~032-0230 (RF $>$ 0.45) and PDF No.~041-0833 (XCAPE candidate less stable than the reference structure), yielding 13 cases in total.

As shown in the left panel of Figure~\ref{fgr:r2scan}b, the two XCAPE protocols yielded candidates with comparable stability for several cases, but the r$^{2}$SCAN-fidelity searches generally produced candidates with lower $E_{\mathrm{hull, MP}}$. Candidates for PDF Nos.~026-0326, 045-1112, and 041-0833 were located within 10~meV/atom of the convex hull under both protocols, and the candidate for PDF No.~041-0833 converged to a structure equivalent to the reference structure after geometry relaxation. Two additional cases fell below this threshold only when the r$^{2}$SCAN-fidelity model was used. One of them had an RF exceeding 0.45, whereas the candidate for PDF No.~062-0286 showed close agreement with the experimental PXRD pattern (RF $=$ 0.05), as shown in the middle and right panels of Figure~\ref{fgr:r2scan}b. These results suggest that evaluating candidates across multiple MLIP fidelities, while retaining PXRD agreement as an independent criterion, can improve the identification of thermodynamically competitive structure assignments from experimental PXRD data.

We selected 12 promising candidates for validation using DFT calculations: three identified through r$^{2}$SCAN-fidelity calculations and nine identified exclusively through PBE(+U)-fidelity calculations. For all candidates, additional geometry relaxations were performed starting from the MLIP-relaxed structures, following the computational protocols used to construct the MP database. Because MP is recomputing all Yb compounds with an updated pseudopotential~\cite{mp_yb_psp}, the competing phases for the three Yb-containing candidates (PDF Nos.~027-0458, 042-1004, and 045-1112) were treated separately (see Methods). As shown in Figure~S14, the MLIP predictions of $E_{\mathrm{hull, MP}}$ agreed with the DFT results to within 12~meV/atom, except for two Ce-containing compounds. Although SevenNet-Omni tended to overstabilize these compounds, DFT calculations confirmed that \ch{CeMoF7} (PDF No.~038-1497) remained below the MP convex hull, while \ch{NaCeS3} (PDF No.~051-0972) exhibited a low $E_{\mathrm{hull, MP}}$ of 17.9~meV/atom. Consequently, 10 of the 12 candidates evaluated by DFT lie within 10~meV/atom of the MP convex hull, the exceptions being \ch{NaCeS3} and \ch{K3YbF6} (17.5~meV/atom).

\section{Discussion}

Several considerations are important when interpreting the structures proposed by XCAPE. The current framework searches ordered structural models in which each Wyckoff site is fully occupied by a single element and therefore does not explicitly represent partial occupancies or substitutional disorder. For experimentally disordered materials, a predicted structure may thus provide an ordered approximation rather than a complete description of the measured phase~\cite{leeman_challenges}. For PDF No.~023-1061 (\ch{CrMo3S4}), for example, XCAPE identified a structure with the same structural framework as the ICSD reference (ICSD 40697~\cite{icsd_40697}), which has a partially occupied Cr site with an occupancy of 0.86 and a reported composition of \ch{Cr_{1.73}Mo6S8} (see Figure~S15a). Full occupancy of this site in the XCAPE candidate results in a compositional difference from the reference, which may contribute to the candidate's relatively high $E_{\mathrm{hull, MP}}$ of 41.2~meV/atom at the r$^{2}$SCAN fidelity.

Energetic ranking also requires careful interpretation. The lowest-energy candidate identified during a search need not correspond to the experimentally observed structure. Metastable phases can persist under particular synthesis conditions~\cite{sun_metastability}, finite-temperature contributions can alter relative thermodynamic stability~\cite{fe_stability}, and MLIP errors can affect the predicted energy ordering~\cite{matbench_discovery}. For PDF No.~055-0103 (\ch{Ta2Zn4O9}), the stability-prioritized candidate reproduced the structure reported by Griesemer et al.~\cite{xrd_prototype}, with an RF of 0.16. The diffraction-prioritized candidate showed substantially closer agreement with the experimental pattern (RF $=$ 0.02), while its $E_{\mathrm{hull, MP}}$ was only 6.6~meV/atom higher at the PBE(+U) fidelity (see Figure~S15b). This energy difference is comparable in magnitude to the deviations between MLIP- and DFT-calculated hull energies observed for several candidates (see Figure~S14), suggesting that the predicted energy ordering alone does not provide a decisive basis for distinguishing these two structures. Calculated energies should therefore be treated as complementary evidence for evaluating diffraction-compatible candidates, rather than as a definitive criterion for structural assignment.

Further experimental validation is needed to establish the proposed structural assignments. The PDF entries considered here provide peak positions and relative intensities but lack the full diffraction profiles required for Rietveld refinement~\cite{rietveld}. Specifically, these peak lists do not preserve the measured intensity at each $2\theta$ step, including the background and peak shapes, and therefore do not support point-by-point profile fitting. Access to the original diffraction profiles or new diffraction measurements would enable Rietveld refinement using the XCAPE candidates as starting models.

\section{Conclusion}

In this work, we developed XCAPE, a PXRD-constrained CSP framework for resolving crystal structures from diffraction data lacking atomic coordinates. XCAPE combines symmetry constraints, evolutionary programming with symmetry-aware mutation operations, and geometry relaxation using a universal MLIP. This integration enables efficient exploration of symmetry-compatible configurations while simultaneously considering energetic plausibility and agreement with the experimental PXRD pattern. XCAPE recovered 71 of 77 known structures (92.2\%) from simulated benchmark patterns and all 12 reference structures from experimental PXRD patterns. When applied to PDF entries previously examined using a prototype-based approach, XCAPE reproduced many reported assignments and frequently identified candidates with lower $E_{\mathrm{hull, MP}}$ and comparable or lower RF values than the previously reported structures. Cross-fidelity analysis further showed that MLIP fidelity can influence both candidate stability and search outcomes. Overall, XCAPE yielded 12 candidate structure assignments consistent with the experimental PXRD patterns but without matching structures in the ICSD, and DFT calculations showed that 10 of these 12 candidates lie within 10~meV/atom of the MP convex hull. These results demonstrate that combining symmetry-constrained global optimization with universal MLIPs provides an efficient approach for extracting plausible atomic structures from underdetermined experimental PXRD data. As autonomous synthesis and high-throughput characterization platforms~\cite{alab,leeman_challenges} generate increasing numbers of materials with unresolved diffraction patterns, XCAPE could facilitate their rapid structural assignment and accelerate the identification of previously uncharacterized crystalline phases.

\section{Methods}

\subsection{Computational framework and structural analysis}
XCAPE was implemented in Python to provide a unified workflow compatible with MLIPs. Structures were generated and modified using the PyXtal library (version 1.1.2)~\cite{pyxtal}. The Atomic Simulation Environment (ASE)~\cite{ase} served as the interface to the MLIP, and PXRD patterns were calculated using Pymatgen~\cite{pymatgen}. Crystal structures were visualized using VESTA~\cite{vesta}. The RF was calculated following the definition of Griesemer et al.~\cite{xrd_prototype}, which is implemented in the MINT package~\cite{mint}, with a wider peak-pairing tolerance. First, each calculated peak was paired with the nearest observed peak within a $2\theta$ tolerance of 0.3$^\circ$. When multiple calculated peaks were assigned to the same observed peak, as can occur for overlapping reflections, their intensities were summed. Calculated peaks without an observed counterpart were retained by assigning $I_i^{\mathrm{obs}}=0$, whereas unmatched observed peaks were retained by assigning $I_i^{\mathrm{calc}}=0$. Thus, both spurious calculated reflections and missing experimental reflections contributed to the residual. After peak pairing, the calculated intensities were multiplied by a single least-squares scaling factor ($k$), while the observed intensities were left unchanged. The scaling factor was obtained by minimizing
\begin{equation}
\begin{split}
\sum_i \left(k I_i^{\mathrm{calc}}-I_i^{\mathrm{obs}}\right)^2
\end{split}
\end{equation}
which gives
\begin{equation}
\begin{split}
k=\frac{\sum_i I_i^{\mathrm{calc}}I_i^{\mathrm{obs}}}{\sum_i \left(I_i^{\mathrm{calc}}\right)^2}
\end{split}
\label{eq:k}
\end{equation}
 The RF was then calculated as follows: \begin{equation}
\begin{split}
\mathrm{RF}=\frac{\sum_i\left(kI_i^{\mathrm{calc}}-I_i^{\mathrm{obs}}\right)^2}{\sum_i\left(I_i^{\mathrm{obs}}\right)^2}
\end{split}
\label{eq:RF}
\end{equation}
Least-squares scaling was used throughout this work, although normalization by either the maximum intensity or the sum of intensities is also available in XCAPE as an optional setting. All PXRD patterns were calculated using CuK$\alpha$ radiation over a $2\theta$ range of 1$^\circ$--90$^\circ$.

\subsection{Evolutionary-programming-based search algorithm}

We adopted an evolutionary programming algorithm~\cite{fogel_ep} in which new candidates are generated exclusively by mutation. Crossover, commonly used in genetic algorithms~\cite{deaven_ga,uspex}, was excluded because it can break the space group symmetry imposed throughout the search. The mutation operations are described in Section~\ref{sec:overview}, and their default probabilities were set as follows: swap same multiplicity (0.25), split (0.1), merge (0.1), swap different element (0.15), reconfigure (0.2), resample (0.1), and multi-resample (0.1). The population size was set to 50 structures per generation. All structures in the first generation were randomly generated. In each subsequent generation, 10\% were randomly generated and the remainder were obtained by mutating parents. In each generation, the parent pool was updated from the newly relaxed structures together with the parent pool of the previous generation in three steps. First, structures were retained only if their energy or RF was within a prescribed margin of the lowest value among these candidates (0.1~eV/atom and 0.2, respectively). Second, duplicate structures were removed. Two structures were regarded as duplicates if their energies and RF values differed by no more than 5~meV/atom and 0.05, respectively, and they were matched by the \textsf{StructureMatcher} module in Pymatgen~\cite{pymatgen} using the tolerances described in the structure recovery criterion. Third, if the number of retained structures did not exceed the maximum number of parents (25), all were used as parents. Otherwise, half of the parents were selected in ascending order of energy and the other half in ascending order of RF. For each mutation, a parent was drawn from this pool by Boltzmann sampling with weights proportional to $\exp[-(E_i-E_{\min})/T]$, where $T = 1.0\times0.9^{\,(g-1)}$~eV/atom in generation $g$, with a minimum of 0.5~eV/atom. To maintain diversity, a parent was instead chosen uniformly at random with a probability of 0.2. For comparison, two alternative search strategies were tested. In the random-only search, all candidate structures were generated independently and relaxed using the universal MLIP. In S\_XCAPE, 70\% of each population was generated randomly, following the original SPINNER implementation~\cite{spinner}. Because the unit cell parameters were fixed to the experimental values, the lattice mutation operator of SPINNER was not applicable. Its 10\% share was therefore assigned to permutation, yielding a permutation probability of 0.3. Pseudocode for XCAPE is provided in Note~S1.

\subsection{Structure recovery criterion}

Structure recovery was assessed using the \textsf{StructureMatcher} module in Pymatgen~\cite{pymatgen} with a fractional length tolerance of 0.2, a site tolerance of 0.3, and an angle tolerance of 5$^\circ$, without reduction to the primitive cell. A reference structure was considered recovered when an XCAPE candidate matched it under these tolerances, and the generation at which this first occurred was reported as the number of generations required for recovery. All matches were further confirmed by manual inspection of the crystal structures. The same criterion was applied to both the simulated benchmark and the experimental validation sets.

\subsection{Machine-learning interatomic potential and geometry optimization}\label{sec:mlip}

The SevenNet-Omni~\cite{7net_omni,sevennet} model was used throughout this work. Unless otherwise specified, the \textsf{mpa} channel of SevenNet-Omni, trained on PBE(+U)-level DFT calculations, was used. During XCAPE runs, only atomic positions were relaxed using the FIRE algorithm~\cite{fire} with a force convergence criterion of 0.02~eV/\AA. Relaxation proceeded in two stages. In the first stage, up to $5N$ steps were performed, where $N$ is the number of atoms in the unit cell, and structures with any interatomic distance shorter than the corresponding distance-matrix value were discarded to exclude unphysical configurations (e.g., collapsed bonds). Details of the distance matrix are provided in Note~S1. Structures that passed this screening were relaxed for up to $20N$ additional steps, and structures that did not converge within these steps were discarded. Unless otherwise specified, all structures were relaxed under symmetry constraints imposed by the \textsf{FixSymmetry} class in ASE~\cite{ase} or TorchSim~\cite{torchsim}. For runs using the hybrid relaxation scheme, the most stable 30\% of the structures in each generation were additionally relaxed without symmetry constraints, starting from their unrelaxed structures. These structures were stored separately and were not included in the parent pool. However, their energies and RF values were evaluated, and they were considered together with the symmetry-constrained structures when the stability-prioritized and diffraction-prioritized solutions were selected. By default, batch relaxations were performed using the TorchSim~\cite{torchsim} package to make efficient use of GPU resources.

\subsection{Stability calculations}

Thermodynamic stability was evaluated using $E_{\mathrm{hull, MP}}$. Because optimized lattice parameters and total energies depend on the computational method, both the target structure and all competing phases in the corresponding chemical system were recalculated using the same MLIP channel and relaxation protocol. All structures were fully relaxed, including both the lattice parameters and atomic coordinates, while their space group symmetry was preserved with \textsf{FixSymmetry} using a symmetry tolerance of $10^{-5}$~\AA. Candidate competing-phase structures were retrieved from the MP database~\cite{mp_db}. When the \textsf{mpa} channel was used, the resulting raw energies were subsequently processed using the Materials Project 2020 (MP2020) compatibility scheme~\cite{mp2020}, including the anion and GGA/GGA+U mixing corrections~\cite{jain_gga_u}, before construction of the phase diagram~\cite{ong_pd} and calculation of $E_{\mathrm{hull, MP}}$.

\subsection{DFT calculations}

DFT calculations were performed using the Vienna Ab initio Simulation Package (VASP)~\cite{vasp_1, vasp_2, vasp_3} with the projector augmented-wave (PAW) method~\cite{blochl_paw,paw}, following the protocols used to construct the MP database~\cite{mp_db}. Input files were generated using the \textsf{MPRelaxSet} and \textsf{MPScanRelaxSet} classes in Pymatgen~\cite{pymatgen}. For candidates identified through PBE(+U)-fidelity calculations, the PBE functional~\cite{pbegga} was used with Hubbard $U$ corrections applied according to the MP convention~\cite{jain_gga_u} and a plane-wave cutoff energy of 520~eV, and the resulting energies were processed using the MP2020 compatibility scheme. For candidates identified through r$^{2}$SCAN-fidelity calculations, the r$^{2}$SCAN functional~\cite{r2scan} was used with a plane-wave cutoff energy of 680~eV. In both cases, the lattice parameters and atomic coordinates were fully relaxed starting from the MLIP-relaxed structures, and each relaxation was repeated once from the resulting structure. The repeated relaxation used the same settings, except that the $k$-point spacing and smearing in the r$^{2}$SCAN calculations were set according to the band gap obtained from the first relaxation, as prescribed by \textsf{MPScanRelaxSet}. $E_{\mathrm{hull, MP}}$ was evaluated using the competing phases in the MP database~\cite{mp_db} calculated at the same level of theory. For \ch{K3YbF6} and \ch{KYbCl3}, the Yb-containing competing phases were taken from the previous \texttt{Yb\_2}-based GGA calculations in the MP database~\cite{mp_db}, which are consistent with the training data of the \textsf{mpa} channel. These phases were combined with the current MP GGA/GGA+U entries, and the MP2020 corrections were applied. For \ch{YbTlI6}, the candidate and all Yb-containing competing phases were relaxed using the MatPES r$^{2}$SCAN settings~\cite{matpes} on which the MLIP was trained, including the PBE\_64 PAW potentials with \texttt{Yb\_h}, a plane-wave cutoff energy of 680~eV, and a $k$-point spacing of 0.22~\AA$^{-1}$. The MP r$^{2}$SCAN~\cite{mp_r2scan} energies were used for the Tl--I phases, which were computed with identical PAW potentials.

\bmsubsection*{Data Availability Statement}

The XCAPE input files, candidate structures, and summary data supporting this study are available on figshare at \url{https://doi.org/10.6084/m9.figshare.34981633}. ICDD PDF entries and ICSD structures are not redistributed owing to license restrictions, but their identifiers are provided in the Supporting Information and the figshare repository.

\bmsubsection*{Code Availability}

The XCAPE source code is available at \url{https://github.com/yslim1/XCAPE}.

\bmsubsection*{Acknowledgments}

This work was supported by the Virtual Engineering Platform Project of the Ministry of Trade, Industry and Energy (MOTIE) of Korea (grant number: P0022336) and by the Nano \& Material Technology Development Program through the National Research Foundation of Korea (NRF) funded by the Ministry of Science and ICT (RS-2026-25542918). The computations were carried out at the Korea Institute of Science and Technology Information (KISTI) National Supercomputing Center (KSC-2025-CRE-0284).

\bmsubsection*{Author Contributions}

Y.L. and S.Ha. conceived the idea. Y.L. and S.Hw. developed the codebase. Y.L. performed the calculations and analyzed the results. Y.L. and S.Ha. prepared the manuscript. All authors contributed to discussions and approved the paper.

\bmsubsection*{Conflict of Interest}

The authors declare no conflict of interest.

\bibliography{refs}

\begin{thebibliography}{}
\providecommand{\doi}[1]{\url{https://doi.org/#1}}
\bibcommenthead

\bibitem[\protect\citeauthoryear{Oganov, Pickard, Zhu, and Needs}{Oganov
  et~al.}{2019}]{structure_disc_1}
Oganov, A.R., C.J. Pickard, Q.~Zhu, and R.J. Needs. 2019.
\newblock Structure prediction drives materials discovery.
\newblock {\em Nat. Rev. Mater.\/}~{\em 4\/}(5): 331--348.
\newblock \doi{10.1038/s41578-019-0101-8} .

\bibitem[\protect\citeauthoryear{Woodley and Catlow}{Woodley and
  Catlow}{2008}]{csp}
Woodley, S.M. and R.~Catlow. 2008.
\newblock Crystal structure prediction from first principles.
\newblock {\em Nat. Mater.\/}~{\em 7\/}(12): 937--946.
\newblock \doi{10.1038/nmat2321} .

\bibitem[\protect\citeauthoryear{van~de Walle}{van~de
  Walle}{2008}]{structure_prop_1}
van~de Walle, A. 2008.
\newblock A complete representation of structure--property relationships in
  crystals.
\newblock {\em Nat. Mater.\/}~{\em 7\/}(6): 455--458.
\newblock \doi{10.1038/nmat2200} .

\bibitem[\protect\citeauthoryear{Hautier, Fischer, Ehrlacher, Jain, and
  Ceder}{Hautier et~al.}{2011}]{elem_substitution}
Hautier, G., C.~Fischer, V.~Ehrlacher, A.~Jain, and G.~Ceder. 2011.
\newblock Data mined ionic substitutions for the discovery of new compounds.
\newblock {\em Inorg. Chem.\/}~{\em 50\/}(2): 656--663.
\newblock \doi{10.1021/ic102031h} .

\bibitem[\protect\citeauthoryear{Freysoldt, Grabowski, Hickel, Neugebauer,
  Kresse, Janotti, and Van~de Walle}{Freysoldt et~al.}{2014}]{defect}
Freysoldt, C., B.~Grabowski, T.~Hickel, J.~Neugebauer, G.~Kresse, A.~Janotti,
  and C.G. Van~de Walle. 2014.
\newblock First-principles calculations for point defects in solids.
\newblock {\em Rev. Mod. Phys.\/}~{\em 86\/}(1): 253--305.
\newblock \doi{10.1103/RevModPhys.86.253} .

\bibitem[\protect\citeauthoryear{Rost, Sachet, Borman, Moballegh, Dickey, Hou,
  Jones, Curtarolo, and Maria}{Rost et~al.}{2015}]{entropy_stabilized}
Rost, C.M., E.~Sachet, T.~Borman, A.~Moballegh, E.C. Dickey, D.~Hou, J.L.
  Jones, S.~Curtarolo, and J.P. Maria. 2015.
\newblock Entropy-stabilized oxides.
\newblock {\em Nat. Commun.\/}~6: 8485.
\newblock \doi{10.1038/ncomms9485} .

\bibitem[\protect\citeauthoryear{Pecharsky and Zavalij}{Pecharsky and
  Zavalij}{2003}]{pxrd}
Pecharsky, V.K. and P.Y. Zavalij. 2003.
\newblock {\em Fundamentals of Powder Diffraction and Structural
  Characterization of Materials\/} (1st ed.).
\newblock Boston: Kluwer Academic Publishers.

\bibitem[\protect\citeauthoryear{Harris, Tremayne, and Kariuki}{Harris
  et~al.}{2001}]{pxrd_structure_1}
Harris, K.D.M., M.~Tremayne, and B.M. Kariuki. 2001.
\newblock Contemporary advances in the use of powder {X}-ray diffraction for
  structure determination.
\newblock {\em Angew. Chem. Int. Ed.\/}~{\em 40\/}(9): 1626--1651.
\newblock \doi{10.1002/1521-3773(20010504)40:9<1626::AID-ANIE16260>3.0.CO;2-7}
  .

\bibitem[\protect\citeauthoryear{{\v{C}}ern{\'y} and
  Favre-Nicolin}{{\v{C}}ern{\'y} and Favre-Nicolin}{2007}]{pxrd_structure_2}
{\v{C}}ern{\'y}, R. and V.~Favre-Nicolin. 2007.
\newblock Direct space methods of structure determination from powder
  diffraction: principles, guidelines and perspectives.
\newblock {\em Z. Kristallogr. Cryst. Mater.\/}~{\em 222\/}(3-4): 105--113.
\newblock \doi{10.1524/zkri.2007.222.3-4.105} .

\bibitem[\protect\citeauthoryear{David and Shankland}{David and
  Shankland}{2008}]{pxrd_hard_1}
David, W.I.F. and K.~Shankland. 2008.
\newblock Structure determination from powder diffraction data.
\newblock {\em Acta Crystallogr. A\/}~{\em 64\/}(1): 52--64.
\newblock \doi{10.1107/S0108767307064252} .

\bibitem[\protect\citeauthoryear{Markvardsen, David, Johnson, and
  Shankland}{Markvardsen et~al.}{2001}]{pxrd_hard_2}
Markvardsen, A.J., W.I.F. David, J.C. Johnson, and K.~Shankland. 2001.
\newblock A probabilistic approach to space-group determination from powder
  diffraction data.
\newblock {\em Acta Crystallogr. A\/}~{\em 57\/}(1): 47--54.
\newblock \doi{10.1107/S0108767300012174} .

\bibitem[\protect\citeauthoryear{Racioppi, Otero-de-la Roza, Hajinazar, and
  Zurek}{Racioppi et~al.}{2025}]{pxrd_hard_3}
Racioppi, S., A.~Otero-de-la Roza, S.~Hajinazar, and E.~Zurek. 2025.
\newblock Powder {X}-ray diffraction assisted evolutionary algorithm for
  crystal structure prediction.
\newblock {\em Digital Discovery\/}~{\em 4\/}(1): 73--83.
\newblock \doi{10.1039/d4dd00269e} .

\bibitem[\protect\citeauthoryear{Leeman, Liu, Stiles, Lee, Bhatt, Schoop, and
  Palgrave}{Leeman et~al.}{2024}]{leeman_challenges}
Leeman, J., Y.~Liu, J.~Stiles, S.B. Lee, P.~Bhatt, L.M. Schoop, and R.G.
  Palgrave. 2024.
\newblock Challenges in high-throughput inorganic materials prediction and
  autonomous synthesis.
\newblock {\em PRX Energy\/}~{\em 3\/}(1): 011002.
\newblock \doi{10.1103/PRXEnergy.3.011002} .

\bibitem[\protect\citeauthoryear{Kabekkodu, Dosen, and Blanton}{Kabekkodu
  et~al.}{2024}]{icdd_5}
Kabekkodu, S.N., A.~Dosen, and T.N. Blanton. 2024.
\newblock {PDF-5+}: a comprehensive {Powder Diffraction File}{\texttrademark}
  for materials characterization.
\newblock {\em Powder Diffr.\/}~{\em 39\/}(2): 47--59.
\newblock \doi{10.1017/S0885715624000150} .

\bibitem[\protect\citeauthoryear{Oganov and Glass}{Oganov and
  Glass}{2006}]{uspex}
Oganov, A.R. and C.W. Glass. 2006.
\newblock Crystal structure prediction using ab initio evolutionary techniques:
  Principles and applications.
\newblock {\em J. Chem. Phys.\/}~{\em 124\/}(24): 244704.
\newblock \doi{10.1063/1.2210932} .

\bibitem[\protect\citeauthoryear{Kang, Jeong, Hong, Hwang, Yoon, and Han}{Kang
  et~al.}{2022}]{spinner}
Kang, S., W.~Jeong, C.~Hong, S.~Hwang, Y.~Yoon, and S.~Han. 2022.
\newblock Accelerated identification of equilibrium structures of
  multicomponent inorganic crystals using machine learning potentials.
\newblock {\em npj Comput. Mater.\/}~{\em 8\/}(1): 108.
\newblock \doi{10.1038/s41524-022-00792-w} .

\bibitem[\protect\citeauthoryear{Griesemer, Ward, and Wolverton}{Griesemer
  et~al.}{2021}]{xrd_prototype}
Griesemer, S.D., L.~Ward, and C.~Wolverton. 2021.
\newblock High-throughput crystal structure solution using prototypes.
\newblock {\em Phys. Rev. Mater.\/}~{\em 5\/}(10): 105003.
\newblock \doi{10.1103/PhysRevMaterials.5.105003} .

\bibitem[\protect\citeauthoryear{Meredig and Wolverton}{Meredig and
  Wolverton}{2013}]{fpass_1}
Meredig, B. and C.~Wolverton. 2013.
\newblock A hybrid computational--experimental approach for automated crystal
  structure solution.
\newblock {\em Nat. Mater.\/}~{\em 12\/}(2): 123--127.
\newblock \doi{10.1038/nmat3490} .

\bibitem[\protect\citeauthoryear{Ward, Michel, and Wolverton}{Ward
  et~al.}{2017}]{fpass_2}
Ward, L., K.~Michel, and C.~Wolverton. 2017.
\newblock Automated crystal structure solution from powder diffraction data:
  Validation of the first-principles-assisted structure solution method.
\newblock {\em Phys. Rev. Mater.\/}~{\em 1\/}(6): 063802.
\newblock \doi{10.1103/PhysRevMaterials.1.063802} .

\bibitem[\protect\citeauthoryear{Saal, Kirklin, Aykol, Meredig, and
  Wolverton}{Saal et~al.}{2013}]{oqmd}
Saal, J.E., S.~Kirklin, M.~Aykol, B.~Meredig, and C.~Wolverton. 2013.
\newblock Materials design and discovery with high-throughput density
  functional theory: {The Open Quantum Materials Database (OQMD)}.
\newblock {\em JOM\/}~{\em 65\/}(11): 1501--1509.
\newblock \doi{10.1007/s11837-013-0755-4} .

\bibitem[\protect\citeauthoryear{Parackal, Goodall, Faber, and
  Armiento}{Parackal et~al.}{2024}]{wyckoff_prototype}
Parackal, A.S., R.E.A. Goodall, F.A. Faber, and R.~Armiento. 2024.
\newblock Identifying crystal structures beyond known prototypes from {X}-ray
  powder diffraction spectra.
\newblock {\em Phys. Rev. Mater.\/}~{\em 8\/}(10): 103801.
\newblock \doi{10.1103/PhysRevMaterials.8.103801} .

\bibitem[\protect\citeauthoryear{Lee, Oba, Ohba, and Kajita}{Lee
  et~al.}{2023}]{evolvmorph}
Lee, J., J.~Oba, N.~Ohba, and S.~Kajita. 2023.
\newblock Creation of crystal structure reproducing {X}-ray diffraction pattern
  without using database.
\newblock {\em npj Comput. Mater.\/}~{\em 9\/}(1): 135.
\newblock \doi{10.1038/s41524-023-01096-3} .

\bibitem[\protect\citeauthoryear{Kim, You, Park, Lim, Kang, Kim, Jeon, Ju,
  Hong, Lee, et~al.}{Kim et~al.}{2026}]{7net_omni}
Kim, J., J.~You, Y.~Park, Y.~Lim, Y.~Kang, J.~Kim, H.~Jeon, S.~Ju, D.~Hong,
  S.Y. Lee, et~al. 2026.
\newblock Optimizing cross-domain transfer for universal machine learning
  interatomic potentials.
\newblock {\em Nat. Commun.\/}~{\em 17\/}(1): 3432.
\newblock \doi{10.1038/s41467-026-70195-8} .

\bibitem[\protect\citeauthoryear{Park, Kim, Hwang, and Han}{Park
  et~al.}{2024}]{sevennet}
Park, Y., J.~Kim, S.~Hwang, and S.~Han. 2024.
\newblock Scalable parallel algorithm for graph neural network interatomic
  potentials in molecular dynamics simulations.
\newblock {\em J. Chem. Theory Comput.\/}~{\em 20\/}(11): 4857--4868.
\newblock \doi{10.1021/acs.jctc.4c00190} .

\bibitem[\protect\citeauthoryear{Deringer, Caro, and Cs{\'a}nyi}{Deringer
  et~al.}{2019}]{mlip_1}
Deringer, V.L., M.A. Caro, and G.~Cs{\'a}nyi. 2019.
\newblock Machine learning interatomic potentials as emerging tools for
  materials science.
\newblock {\em Adv. Mater.\/}~{\em 31\/}(46): 1902765.
\newblock \doi{10.1002/adma.201902765} .

\bibitem[\protect\citeauthoryear{Behler}{Behler}{2016}]{mlip_2}
Behler, J. 2016.
\newblock Perspective: Machine learning potentials for atomistic simulations.
\newblock {\em J. Chem. Phys.\/}~{\em 145\/}(17): 170901.
\newblock \doi{10.1063/1.4966192} .

\bibitem[\protect\citeauthoryear{Podryabinkin, Tikhonov, Shapeev, and
  Oganov}{Podryabinkin et~al.}{2019}]{mlip_csp}
Podryabinkin, E.V., E.V. Tikhonov, A.V. Shapeev, and A.R. Oganov. 2019.
\newblock Accelerating crystal structure prediction by machine-learning
  interatomic potentials with active learning.
\newblock {\em Phys. Rev. B\/}~{\em 99\/}(6): 064114.
\newblock \doi{10.1103/PhysRevB.99.064114} .

\bibitem[\protect\citeauthoryear{Jain, Ong, Hautier, Chen, Richards, Dacek,
  Cholia, Gunter, Skinner, Ceder, and Persson}{Jain et~al.}{2013}]{mp_db}
Jain, A., S.P. Ong, G.~Hautier, W.~Chen, W.D. Richards, S.~Dacek, S.~Cholia,
  D.~Gunter, D.~Skinner, G.~Ceder, and K.A. Persson. 2013.
\newblock Commentary: {The Materials Project}: A materials genome approach to
  accelerating materials innovation.
\newblock {\em APL Mater.\/}~{\em 1\/}(1): 011002.
\newblock \doi{10.1063/1.4812323} .

\bibitem[\protect\citeauthoryear{Zagorac, M{\"u}ller, Ruehl, Zagorac, and
  Rehme}{Zagorac et~al.}{2019}]{icsd}
Zagorac, D., H.~M{\"u}ller, S.~Ruehl, J.~Zagorac, and S.~Rehme. 2019.
\newblock Recent developments in the {Inorganic Crystal Structure Database}:
  theoretical crystal structure data and related features.
\newblock {\em J. Appl. Crystallogr.\/}~{\em 52\/}(5): 918--925.
\newblock \doi{10.1107/S160057671900997X} .

\bibitem[\protect\citeauthoryear{Fredericks, Parrish, Sayre, and
  Zhu}{Fredericks et~al.}{2021}]{pyxtal}
Fredericks, S., K.~Parrish, D.~Sayre, and Q.~Zhu. 2021.
\newblock {PyXtal}: A {Python} library for crystal structure generation and
  symmetry analysis.
\newblock {\em Comput. Phys. Commun.\/}~261: 107810.
\newblock \doi{10.1016/j.cpc.2020.107810} .

\bibitem[\protect\citeauthoryear{Hwang, Lee, Kim, Han, Kang, and Kang}{Hwang
  et~al.}{2026}]{topic}
Hwang, S., J.~Lee, J.~Kim, S.~Han, Y.~Kang, and S.~Kang. 2026.
\newblock Discovery of oxide {Li}-conducting electrolytes with corner-sharing
  frameworks via topology-constrained crystal structure prediction.
\newblock {\em J. Mater. Chem. A\/}~{\em 14\/}(33): 21940--21956.
\newblock \doi{10.1039/d6ta01092j} .

\bibitem[\protect\citeauthoryear{Cussen}{Cussen}{2010}]{cussen_garnet}
Cussen, E.J. 2010.
\newblock Structure and ionic conductivity in lithium garnets.
\newblock {\em J. Mater. Chem.\/}~{\em 20\/}(25): 5167--5173.
\newblock \doi{10.1039/b925553b} .

\bibitem[\protect\citeauthoryear{Delgado, Mora, Pérez, and González}{Delgado
  et~al.}{2007}]{icsd_157537}
Delgado, G.E., A.J. Mora, F.V. Pérez, and J.~González. 2007.
\newblock Crystal structure of the ternary semiconductor compound thallium
  gallium sulfide, {TlGaS$_2$}.
\newblock {\em Physica B\/}~{\em 391\/}(2): 385--388.
\newblock \doi{10.1016/j.physb.2006.10.030} .

\bibitem[\protect\citeauthoryear{Badding and DiSalvo}{Badding and
  DiSalvo}{1990}]{icsd_40207}
Badding, M.E. and F.J. DiSalvo. 1990.
\newblock Synthesis and structure of tantalum silicide telluride,
  {Ta$_4$SiTe$_4$}, a new low-dimensional material.
\newblock {\em Inorg. Chem.\/}~{\em 29\/}(20): 3952--3954.
\newblock \doi{10.1021/ic00345a009} .

\bibitem[\protect\citeauthoryear{Buschow, {van Engen}, and {de Mooij}}{Buschow
  et~al.}{1984}]{icsd_104927}
Buschow, K.H.J., P.G. {van Engen}, and D.B. {de Mooij}. 1984.
\newblock Magnetic and magneto-optical properties of {Heusler} alloys of the
  type {Ni$_{3-x}$Mn$_x$Sn}.
\newblock {\em J. Magn. Magn. Mater.\/}~{\em 40\/}(3): 339--347.
\newblock \doi{10.1016/0304-8853(84)90327-5} .

\bibitem[\protect\citeauthoryear{Sun, Dacek, Ong, Hautier, Jain, Richards,
  Gamst, Persson, and Ceder}{Sun et~al.}{2016}]{sun_metastability}
Sun, W., S.T. Dacek, S.P. Ong, G.~Hautier, A.~Jain, W.D. Richards, A.C. Gamst,
  K.A. Persson, and G.~Ceder. 2016.
\newblock The thermodynamic scale of inorganic crystalline metastability.
\newblock {\em Sci. Adv.\/}~{\em 2\/}(11): e1600225.
\newblock \doi{10.1126/sciadv.1600225} .

\bibitem[\protect\citeauthoryear{Perdew, Burke, and Ernzerhof}{Perdew
  et~al.}{1996}]{pbegga}
Perdew, J.P., K.~Burke, and M.~Ernzerhof. 1996.
\newblock Generalized gradient approximation made simple.
\newblock {\em Phys. Rev. Lett.\/}~{\em 77\/}(18): 3865--3868.
\newblock \doi{10.1103/PhysRevLett.77.3865} .

\bibitem[\protect\citeauthoryear{Dudarev, Botton, Savrasov, Humphreys, and
  Sutton}{Dudarev et~al.}{1998}]{dudarev}
Dudarev, S.L., G.A. Botton, S.Y. Savrasov, C.J. Humphreys, and A.P. Sutton.
  1998.
\newblock Electron-energy-loss spectra and the structural stability of nickel
  oxide: {An} {LSDA+U} study.
\newblock {\em Phys. Rev. B\/}~{\em 57\/}(3): 1505--1509.
\newblock \doi{10.1103/PhysRevB.57.1505} .

\bibitem[\protect\citeauthoryear{Jain, Hautier, Ong, Moore, Fischer, Persson,
  and Ceder}{Jain et~al.}{2011}]{jain_gga_u}
Jain, A., G.~Hautier, S.P. Ong, C.J. Moore, C.C. Fischer, K.A. Persson, and
  G.~Ceder. 2011.
\newblock Formation enthalpies by mixing {GGA} and {GGA}+{U} calculations.
\newblock {\em Phys. Rev. B\/}~{\em 84\/}(4): 045115.
\newblock \doi{10.1103/PhysRevB.84.045115} .

\bibitem[\protect\citeauthoryear{Perdew, Ruzsinszky, Csonka, Vydrov, Scuseria,
  Constantin, Zhou, and Burke}{Perdew et~al.}{2008}]{pbe_over}
Perdew, J.P., A.~Ruzsinszky, G.I. Csonka, O.A. Vydrov, G.E. Scuseria, L.A.
  Constantin, X.~Zhou, and K.~Burke. 2008.
\newblock Restoring the density-gradient expansion for exchange in solids and
  surfaces.
\newblock {\em Phys. Rev. Lett.\/}~{\em 100\/}(13): 136406.
\newblock \doi{10.1103/PhysRevLett.100.136406} .

\bibitem[\protect\citeauthoryear{Sun, Ruzsinszky, and Perdew}{Sun
  et~al.}{2015}]{scan}
Sun, J., A.~Ruzsinszky, and J.P. Perdew. 2015.
\newblock Strongly constrained and appropriately normed semilocal density
  functional.
\newblock {\em Phys. Rev. Lett.\/}~{\em 115\/}(3): 036402.
\newblock \doi{10.1103/PhysRevLett.115.036402} .

\bibitem[\protect\citeauthoryear{Furness, Kaplan, Ning, Perdew, and
  Sun}{Furness et~al.}{2020}]{r2scan}
Furness, J.W., A.D. Kaplan, J.~Ning, J.P. Perdew, and J.~Sun. 2020.
\newblock Accurate and numerically efficient {r$^2$SCAN} meta-generalized
  gradient approximation.
\newblock {\em J. Phys. Chem. Lett.\/}~{\em 11\/}(19): 8208--8215.
\newblock \doi{10.1021/acs.jpclett.0c02405} .

\bibitem[\protect\citeauthoryear{Kaplan, Liu, Qi, Ko, Deng, Riebesell, Ceder,
  Persson, and Ong}{Kaplan et~al.}{2025}]{matpes}
Kaplan, A.D., R.~Liu, J.~Qi, T.W. Ko, B.~Deng, J.~Riebesell, G.~Ceder, K.A.
  Persson, and S.P. Ong. 2025.
\newblock A foundational potential energy surface dataset for materials.
\newblock {\em arXiv preprint\/}.
\newblock \doi{10.48550/arXiv.2503.04070}.
\newblock {\href{https://arxiv.org/abs/2503.04070}{{arXiv:2503.04070}}} .

\bibitem[\protect\citeauthoryear{Kothakonda, Kaplan, Isaacs, Bartel, Furness,
  Ning, Wolverton, Perdew, and Sun}{Kothakonda
  et~al.}{2023}]{kothakonda_r2scan}
Kothakonda, M., A.D. Kaplan, E.B. Isaacs, C.J. Bartel, J.W. Furness, J.~Ning,
  C.~Wolverton, J.P. Perdew, and J.~Sun. 2023.
\newblock Testing the {r$^2$SCAN} density functional for the thermodynamic
  stability of solids with and without a van der {Waals} correction.
\newblock {\em ACS Mater. Au\/}~{\em 3\/}(2): 102--111.
\newblock \doi{10.1021/acsmaterialsau.2c00059} .

\bibitem[\protect\citeauthoryear{{Materials Project}}{{Materials
  Project}}{2023}]{mp_yb_psp}
{Materials Project}. 2023.
\newblock Pseudopotentials: {GGA/GGA+U} calculations.
\newblock
  \url{https://docs.materialsproject.org/methodology/materials-methodology/calculation-details/gga+u-calculations/pseudopotentials}.
\newblock Accessed 2026-10-16.

\bibitem[\protect\citeauthoryear{Harbrecht and Mahne}{Harbrecht and
  Mahne}{1992}]{icsd_40697}
Harbrecht, B. and S.~Mahne. 1992.
\newblock The crystal structure of {Cr$_{1.73}$Mo$_6$S$_8$}.
\newblock {\em J. Alloys Compd.\/}~{\em 178\/}(1--2): 467--475.
\newblock \doi{10.1016/0925-8388(92)90288-K} .

\bibitem[\protect\citeauthoryear{Tolborg, Klarbring, Ganose, and Walsh}{Tolborg
  et~al.}{2022}]{fe_stability}
Tolborg, K., J.~Klarbring, A.M. Ganose, and A.~Walsh. 2022.
\newblock Free energy predictions for crystal stability and synthesisability.
\newblock {\em Digital Discovery\/}~{\em 1\/}(5): 586--595.
\newblock \doi{10.1039/d2dd00050d} .

\bibitem[\protect\citeauthoryear{Riebesell, Goodall, Benner, Chiang, Deng,
  Ceder, Asta, Lee, Jain, and Persson}{Riebesell
  et~al.}{2025}]{matbench_discovery}
Riebesell, J., R.E.A. Goodall, P.~Benner, Y.~Chiang, B.~Deng, G.~Ceder,
  M.~Asta, A.A. Lee, A.~Jain, and K.A. Persson. 2025.
\newblock A framework to evaluate machine learning crystal stability
  predictions.
\newblock {\em Nat. Mach. Intell.\/}~{\em 7\/}(6): 836--847.
\newblock \doi{10.1038/s42256-025-01055-1} .

\bibitem[\protect\citeauthoryear{McCusker, Von~Dreele, Cox, Lou{\"e}r, and
  Scardi}{McCusker et~al.}{1999}]{rietveld}
McCusker, L.B., R.B. Von~Dreele, D.E. Cox, D.~Lou{\"e}r, and P.~Scardi. 1999.
\newblock Rietveld refinement guidelines.
\newblock {\em J. Appl. Crystallogr.\/}~{\em 32\/}(1): 36--50.
\newblock \doi{10.1107/S0021889898009856} .

\bibitem[\protect\citeauthoryear{Szymanski, Rendy, Fei, Kumar, He, Milsted,
  McDermott, Gallant, Cubuk, Merchant, Kim, Jain, Bartel, Persson, Zeng, and
  Ceder}{Szymanski et~al.}{2023}]{alab}
Szymanski, N.J., B.~Rendy, Y.~Fei, R.E. Kumar, T.~He, D.~Milsted, M.J.
  McDermott, M.~Gallant, E.D. Cubuk, A.~Merchant, H.~Kim, A.~Jain, C.J. Bartel,
  K.~Persson, Y.~Zeng, and G.~Ceder. 2023.
\newblock An autonomous laboratory for the accelerated synthesis of inorganic
  materials.
\newblock {\em Nature\/}~{\em 624\/}(7990): 86--91.
\newblock \doi{10.1038/s41586-023-06734-w} .

\bibitem[\protect\citeauthoryear{Hjorth~Larsen, Mortensen, Blomqvist, Castelli,
  Christensen, Du{\l}ak, Friis, Groves, Hammer, Hargus, et~al.}{Hjorth~Larsen
  et~al.}{2017}]{ase}
Hjorth~Larsen, A., J.J. Mortensen, J.~Blomqvist, I.E. Castelli, R.~Christensen,
  M.~Du{\l}ak, J.~Friis, M.N. Groves, B.~Hammer, C.~Hargus, et~al. 2017.
\newblock The atomic simulation environment---a {Python} library for working
  with atoms.
\newblock {\em J. Phys.: Condens. Matter\/}~{\em 29\/}(27): 273002.
\newblock \doi{10.1088/1361-648X/aa680e} .

\bibitem[\protect\citeauthoryear{Ong, Richards, Jain, Hautier, Kocher, Cholia,
  Gunter, Chevrier, Persson, and Ceder}{Ong et~al.}{2013}]{pymatgen}
Ong, S.P., W.D. Richards, A.~Jain, G.~Hautier, M.~Kocher, S.~Cholia, D.~Gunter,
  V.L. Chevrier, K.A. Persson, and G.~Ceder. 2013.
\newblock {Python Materials Genomics (pymatgen)}: A robust, open-source
  {Python} library for materials analysis.
\newblock {\em Comput. Mater. Sci.\/}~68: 314--319.
\newblock \doi{10.1016/j.commatsci.2012.10.028} .

\bibitem[\protect\citeauthoryear{Momma and Izumi}{Momma and
  Izumi}{2011}]{vesta}
Momma, K. and F.~Izumi. 2011.
\newblock {VESTA 3} for three-dimensional visualization of crystal, volumetric
  and morphology data.
\newblock {\em J. Appl. Crystallogr.\/}~{\em 44\/}(6): 1272--1276.
\newblock \doi{10.1107/S0021889811038970} .

\bibitem[\protect\citeauthoryear{Ward and Michel}{Ward and Michel}{2017}]{mint}
Ward, L. and K.~Michel. 2017.
\newblock {MINT}.
\newblock \url{https://github.com/materials/mint}.

\bibitem[\protect\citeauthoryear{Fogel}{Fogel}{1994}]{fogel_ep}
Fogel, D.B. 1994.
\newblock An introduction to simulated evolutionary optimization.
\newblock {\em IEEE Trans. Neural Netw.\/}~{\em 5\/}(1): 3--14.
\newblock \doi{10.1109/72.265956} .

\bibitem[\protect\citeauthoryear{Deaven and Ho}{Deaven and
  Ho}{1995}]{deaven_ga}
Deaven, D.M. and K.M. Ho. 1995.
\newblock Molecular geometry optimization with a genetic algorithm.
\newblock {\em Phys. Rev. Lett.\/}~{\em 75\/}(2): 288--291.
\newblock \doi{10.1103/PhysRevLett.75.288} .

\bibitem[\protect\citeauthoryear{Bitzek, Koskinen, G{\"a}hler, Moseler, and
  Gumbsch}{Bitzek et~al.}{2006}]{fire}
Bitzek, E., P.~Koskinen, F.~G{\"a}hler, M.~Moseler, and P.~Gumbsch. 2006.
\newblock Structural relaxation made simple.
\newblock {\em Phys. Rev. Lett.\/}~{\em 97\/}(17): 170201.
\newblock \doi{10.1103/PhysRevLett.97.170201} .

\bibitem[\protect\citeauthoryear{Cohen, Riebesell, Goodall, Kolluru, Falletta,
  Krause, Colindres, Ceder, and Gangan}{Cohen et~al.}{2025}]{torchsim}
Cohen, O., J.~Riebesell, R.~Goodall, A.~Kolluru, S.~Falletta, J.~Krause,
  J.~Colindres, G.~Ceder, and A.S. Gangan. 2025.
\newblock {TorchSim}: An efficient atomistic simulation engine in {PyTorch}.
\newblock {\em AI for Science\/}~{\em 1\/}(2): 025003.
\newblock \doi{10.1088/3050-287X/ae1799} .

\bibitem[\protect\citeauthoryear{Wang, Kingsbury, McDermott, Horton, Jain, Ong,
  Dwaraknath, and Persson}{Wang et~al.}{2021}]{mp2020}
Wang, A., R.~Kingsbury, M.~McDermott, M.~Horton, A.~Jain, S.P. Ong,
  S.~Dwaraknath, and K.A. Persson. 2021.
\newblock A framework for quantifying uncertainty in {DFT} energy corrections.
\newblock {\em Sci. Rep.\/}~{\em 11\/}(1): 15496.
\newblock \doi{10.1038/s41598-021-94550-5} .

\bibitem[\protect\citeauthoryear{Ong, Wang, Kang, and Ceder}{Ong
  et~al.}{2008}]{ong_pd}
Ong, S.P., L.~Wang, B.~Kang, and G.~Ceder. 2008.
\newblock {Li--Fe--P--O$_2$} phase diagram from first principles calculations.
\newblock {\em Chem. Mater.\/}~{\em 20\/}(5): 1798--1807.
\newblock \doi{10.1021/cm702327g} .

\bibitem[\protect\citeauthoryear{Kresse and Hafner}{Kresse and
  Hafner}{1993}]{vasp_1}
Kresse, G. and J.~Hafner. 1993.
\newblock Ab initio molecular dynamics for liquid metals.
\newblock {\em Phys. Rev. B\/}~{\em 47\/}(1): 558--561.
\newblock \doi{10.1103/PhysRevB.47.558} .

\bibitem[\protect\citeauthoryear{Kresse and Furthmüller}{Kresse and
  Furthmüller}{1996}]{vasp_2}
Kresse, G. and J.~Furthmüller. 1996.
\newblock Efficiency of ab-initio total energy calculations for metals and
  semiconductors using a plane-wave basis set.
\newblock {\em Comput. Mater. Sci.\/}~{\em 6\/}(1): 15--50.
\newblock \doi{10.1016/0927-0256(96)00008-0} .

\bibitem[\protect\citeauthoryear{Kresse and Furthm{\"u}ller}{Kresse and
  Furthm{\"u}ller}{1996}]{vasp_3}
Kresse, G. and J.~Furthm{\"u}ller. 1996.
\newblock Efficient iterative schemes for ab initio total-energy calculations
  using a plane-wave basis set.
\newblock {\em Phys. Rev. B\/}~{\em 54\/}(16): 11169--11186.
\newblock \doi{10.1103/PhysRevB.54.11169} .

\bibitem[\protect\citeauthoryear{Bl{\"o}chl}{Bl{\"o}chl}{1994}]{blochl_paw}
Bl{\"o}chl, P.E. 1994.
\newblock Projector augmented-wave method.
\newblock {\em Phys. Rev. B\/}~{\em 50\/}(24): 17953--17979.
\newblock \doi{10.1103/PhysRevB.50.17953} .

\bibitem[\protect\citeauthoryear{Kresse and Joubert}{Kresse and
  Joubert}{1999}]{paw}
Kresse, G. and D.~Joubert. 1999.
\newblock From ultrasoft pseudopotentials to the projector augmented-wave
  method.
\newblock {\em Phys. Rev. B\/}~{\em 59\/}(3): 1758--1775.
\newblock \doi{10.1103/PhysRevB.59.1758} .

\bibitem[\protect\citeauthoryear{Kingsbury, Gupta, Bartel, Munro, Dwaraknath,
  Horton, and Persson}{Kingsbury et~al.}{2022}]{mp_r2scan}
Kingsbury, R., A.S. Gupta, C.J. Bartel, J.M. Munro, S.~Dwaraknath, M.~Horton,
  and K.A. Persson. 2022.
\newblock Performance comparison of ${r}^{2}\mathrm{SCAN}$ and {SCAN} {metaGGA}
  density functionals for solid materials via an automated, high-throughput
  computational workflow.
\newblock {\em Phys. Rev. Mater.\/}~{\em 6\/}(1): 013801.
\newblock \doi{10.1103/PhysRevMaterials.6.013801} .

\end{thebibliography}

\bmsubsection*{Supporting Information}

The Supporting Information is provided as a separate document (ancillary file).

\end{document}